\documentclass[letterpaper,twocolumn,10pt]{article}
\usepackage{usenix}

\usepackage{tikz}
\usepackage{amsmath}
\usepackage{amssymb}

\usepackage{filecontents}

\usepackage{xurl}
\usepackage{enumitem}
\usepackage{booktabs}
\usepackage{makecell}
\usepackage{pifont}
\usepackage{multirow}
\usepackage{xcolor}
\usepackage{graphicx}
\usepackage{subcaption}
\definecolor{y}{HTML}{4f772d}
\definecolor{n}{HTML}{9e2a2b}
\begin{document}

\date{}

\title{\Large \bf ProvAgent: Threat Detection Based on Identity-Behavior Binding and Multi-Agent Collaborative Attack Investigation}

\author{
    {\rm Wenhao Yan\textsuperscript{1,2},
    Ning An\textsuperscript{1,2},
    Linxu Li\textsuperscript{1,2},
    Bingsheng Bi\textsuperscript{1,2},
    Bo Jiang\textsuperscript{1,2}}\\
    {\rm Zhigang Lu\textsuperscript{1,2},
    Baoxu Liu\textsuperscript{1,2},
    Junrong Liu\textsuperscript{1,2},
    Cong Dong\textsuperscript{3}} \\
    \textsuperscript{1}\textit{Institute of Information Engineering, Chinese Academy of Sciences}\\
    \textsuperscript{2}\textit{University of Chinese Academy of Sciences}\quad
    \textsuperscript{3}\textit{Zhongguancun Laboratory}
}

\maketitle

\begin{abstract}
Advanced Persistent Threats (APTs) pose critical challenges to modern cybersecurity due to their multi-stage and stealthy nature. While provenance-based detection approaches show promise in capturing causal attack semantics, current threat provenance practices face two paradoxical issues: (1) expert skepticism, where human analysts doubt the capability of traditional detection models to identify complex attacks; and (2) expert dependence, as analysts cannot manually process large-scale raw logs to detect threats without these models. Consequently, collaboration between humans and traditional models remains the prevailing paradigm. However, this renders investigation quality contingent upon human expertise and frequently results in alert fatigue.
To address these challenges, we present ProvAgent, a framework that evolves the threat provenance paradigm from human-model collaboration to a novel collaboration between multi-agent systems and traditional models. ProvAgent leverages the speed and cost-efficiency of traditional models for initial anomaly screening over large-scale logs. By enforcing fine-grained identity-behavior consistency via graph contrastive learning, it profiles entities based on specific attributes to generate high-fidelity alerts. With these alerts serving as investigation entry points, ProvAgent achieves in-depth autonomous investigation through a hypothesis-verification multi-agent framework.
Evaluations with real-world datasets demonstrate that ProvAgent outperforms six state-of-the-art (SOTA) baselines in anomaly detection. Through automated investigation, ProvAgent reconstructs near-complete attack processes at a minimum cost of \$0.06 per day.
\end{abstract}

\section{Introduction}
\label{sec:intro}

Advanced Persistent Threats (APTs) have emerged as a paramount challenge in modern cybersecurity, posing severe risks to large-scale enterprises and governmental defense institutions \cite{alshamrani2019aptsurvey,li2021threat,zipperle2022provenance}. The inherent complexity of these multi-stage campaigns, characterized by strategic evasion and sophisticated tactics, frequently renders traditional detection methodologies ineffective. Consequently, current industrial and academic research has pivoted toward Provenance-based Endpoint Detection and Response (P-EDR) systems \cite{dong2023we,inam2023sok, xu2025deep}. In contrast to conventional EDR solutions that operate on isolated events, provenance-based approaches capture rich causal semantics by modeling dependencies among fundamental system entities such as processes, files, and socket. By formalizing system activities as provenance graphs, P-EDR enables reasoning about long-term, multi-stage attack behaviors. This graph-centric paradigm empowers defenders to reconstruct complete attack narratives and expose stealthy adversarial operations that appear benign in isolation but reveal malicious intent within their causal context.

Existing provenance-graph-based approaches for APT detection can be classified into three categories: 
(1) \textbf{Heuristic-based approaches}\cite{hossain2017sleuth,milajerdi2019holmes,hassan2020tactical,xiong2020conan,hossain2020combating,wang2025incorporating} rely on predefined expert rules or TTP mappings, yet exhibit inherent fragility when confronting zero-day exploits and polymorphic adversarial variants. 
(2) \textbf{Anomaly-based approaches} \cite{akbar2022advanced,kapoor2021prov,log2vec,deeplog,ding2023airtag,manzoor2016fast,han2020unicorn,wang2022threatrace, yang2023prographer,rehman2024flash,cheng2024kairos,jia2024magic,goyal2024r,wu2025brewing,qiao2025slot,bilot2025sometimes} leverage graph neural networks to characterize benign behavioral baselines. Yet, these methods frequently misinterpret legitimate operational heterogeneity as anomalies and yield opaque alerts devoid of contextual explainability. This necessitates labor-intensive manual retrieval across voluminous audit logs to correlate disparate events for alert validation. 
(3) \textbf{LLM-based approaches} \cite{song2024audit,mukherjee2025llm,cheng2025omnisec,zhang2025smartguard,zuo2503knowledge} attempt to incorporate large language models into the provenance analysis pipeline, but most existing frameworks employ LLMs primarily as direct classifiers. This practice incurs prohibitive computational overhead and detection latency while restricting contextual reasoning to localized subgraphs, thereby precluding the synthesis of a global attack narrative across massive audit data.

Although these systems have demonstrated impressive capabilities in threat detection, a significant cognitive gap remains in real-world scenarios. Security Operations Centers (SOCs) still rely heavily on human expertise to decipher complex alerts, as human analysts possess the multi-step retrospective reasoning and causal inference skills that current models lack \cite{soc_study}. Nevertheless, alert fatigue continues to plague operational workflows, forcing analysts to manually validate disparate alerts and incurring prohibitive labor costs. While existing alert correlation techniques \cite{alsaheel2021atlas,zeng2021watson, yan2025sentient,aly2025ocr} attempt to alleviate this burden, they generally lack the capacity for active threat investigation and deep reasoning. As a result, these methods fail to satisfy the complex requirements of automated attack provenance and reconstruction. 

To address these challenges, we present ProvAgent, a novel threat provenance paradigm that synergizes multi-agent systems with traditional models. Designed to simulate the cognitive collaboration mechanisms within a Security Operations Center (SOC), ProvAgent aims to achieve effective threat detection and APT reconstruction. ProvAgent comprises two key components: the Entity-Profile-based Detection (EPD) module and the Multi-Agent Investigation (MAI) module.  EPD performs rapid threat detection from audit logs, prioritizing high-fidelity alerts with minimal false positives. Subsequently, MAI emulates the collaborative dynamics among human experts to conduct in-depth analysis and validation of the threats identified by EPD.

Specifically, during the threat detection phase, EPD employs graph contrastive learning to model the intrinsic relationship between the identities of system entities and their behaviors. It captures consistent behavior patterns exhibited by entities with specific identities, such as processes with particular names, files at specific paths, or sockets bound to specific ports, across different time periods. EPD detects anomalies by contrasting observed behaviors against the learned identity profiles.

During the attack investigation phase, MAI reduces reliance on human analysts with uncertain expertise by introducing large language models that possess robust reasoning capabilities and deep domain knowledge. By orchestrating a team of agents equipped with independent reasoning and collaborative investigation capabilities, MAI achieves iterative reasoning and evidentiary cross-validation. This process operates within a closed-loop "hypothesis-verification" framework, supported by a benign behavior reference system, to reconstruct complete attack steps.

We conducted a comprehensive evaluation on the public datasets from DARPA E3~\cite{darpa_e3}, E5~\cite{darpa_e5}, and OpTC~\cite{darpa_optc}, demonstrating that ProvAgent's anomaly detection capabilities surpass existing state-of-the-art (SOTA) methods. By leveraging alerts generated by EPD as investigative leads, ProvAgent precisely identifies false positives and expands upon established clues to uncover genuine attacks that were initially missed by EPD.

In summary, this paper makes the following contributions:
\begin{itemize}[leftmargin=*,topsep=0pt,itemsep=0pt,partopsep=0pt,parsep=0pt]
    \item We construct a synergistic framework that combines the efficiency and cost-effectiveness of traditional models in alert generation with the robust reasoning capabilities of a multi-agent architecture.
    \item We propose Entity-Profile-based Detection (EPD), which leverages Graph Contrastive Learning to enforce fine-grained identity-behavior binding to generate high-fidelity alerts.
    \item We propose a multi-agent framework (MAI) that leverages the powerful reasoning capabilities of LLMs to conduct in-depth APT analysis, eliminating false positives and uncovering missed detections.
    \item We conducted a comprehensive evaluation using real-world datasets, demonstrating the effectiveness and cost-effectiveness of ProvAgent in detecting APTs.
\end{itemize}

\section{Background}
\subsection{Excessive Alert Noise}
Most existing approaches attempt to reconstruct complete APT attack chains solely through automated detection models. However, achieving comprehensive and accurate APT detection without human-in-the-loop intervention remains nearly impossible with simple models alone~\cite{soc_study}. To demonstrate effectiveness, these methods have become excessively focused on optimizing Recall and F1-score metrics, often incorporating high-volume events such as port scanning during lateral movement as detection targets and characterizing them as APT attacks. While the sheer volume of port scanning events can inflate performance metrics and obscure the actual precision deficits in detecting genuine APT behaviors, this practice fundamentally deviates from the fundamental goals of APT detection.

In reality, these are not the alerts we truly need. Every alert generated by EDR necessitates complex investigation and verification by a professional security analyst. In real-world SOC scenarios, analysts are overwhelmed by this massive volume of low-quality alerts. Therefore, we argue that the core task of EDR is to identify critical attack pivot points with minimal false positives, rather than attempting to uncover the entire attack surface.

\textbf{The lack of Identity-Behavior consistency leads to high false positives.} 
Methods such as ShadeWatcher~\cite{zengy2022shadewatcher}, Kairos~\cite{cheng2024kairos}, and Orthrus~\cite{jiang2025orthrus} employ GNNs and causal analysis to learn the interaction patterns of system entities; however, they overlook the consistency between an entity's behavior and its identity. These approaches primarily focus on capturing topological dependency structures, causing the models to target topological anomalies. Consequently, it becomes difficult to distinguish benign administrative operations from malicious attacks that exhibit similar interaction patterns, resulting in numerous false positives. While prior works such as Threatrace~\cite{wang2022threatrace}, Flash~\cite{rehman2024flash}, and OCR-APT~\cite{aly2025ocr} have noted the importance of Identity-Behavior consistency, they merely correlate behavior with coarse-grained identity information. Specifically, node types such as \textit{file}, \textit{subject}, and \textit{netflow}. 
However, even nodes within the same category may perform completely disparate operations in reality. For instance, while both \texttt{cat} and \texttt{nginx} processes belong to the \textit{subject} category, \texttt{cat} typically involves file reading, whereas \texttt{nginx} involves network operations. While capable of detecting some anomalies, this coarse-grained binding inevitably generates excessive high-confidence false positives.

\subsection{Superficial investigation}
Even when the core pivot points of an APT attack are detected, this remains far from sufficient. Existing methods such as WATSON~\cite{zeng2021watson}, ATLAS~\cite{alsaheel2021atlas}, Flash~\cite{rehman2024flash}, Sentient~\cite{yan2025sentient}, and Orthrus~\cite{jiang2025orthrus} attempt to use graph algorithms to correlate alerts and construct extensive alert graphs. However, these graphs suffer from significant false negatives and contain numerous false positives, implying that SOCs still rely on manual threat investigation and analysis. As previously mentioned, threat investigation and analysis by human experts is an indispensable part of APT attribution due to the highly complex and stealthy nature of these attacks. Nevertheless, since APT attacks are extremely rare events, security analysts often spend prolonged periods repeatedly ruling out false positives, which can, to some extent, lead to alert fatigue and relaxed vigilance.

Employing LLMs, which possess extensive cross-domain knowledge including security expertise~\cite{ullah2024llms,zibaeirad2025reasoning,habibzadeh2025large}, to replace human security analysts in threat investigation and analysis appears to offer a solution to this challenge.
While OCR-APT~\cite{aly2025ocr} attempts to leverage LLM capabilities to correlate alerts and generate human-readable reports, it restricts the model to the role of an expert advisor within a predefined workflow, where it lacks autonomous investigative authority and collaborative investigative capabilities. The efficacy of the reporting mechanism is inherently limited by the granularity of the ingested alerts. In this case, the system exhibited a bias toward noisy, low-level events like port scanning, while simultaneously overlooking the critical stages of the attack.

\begin{figure}
    \centering
    \includegraphics[width=0.85\linewidth]{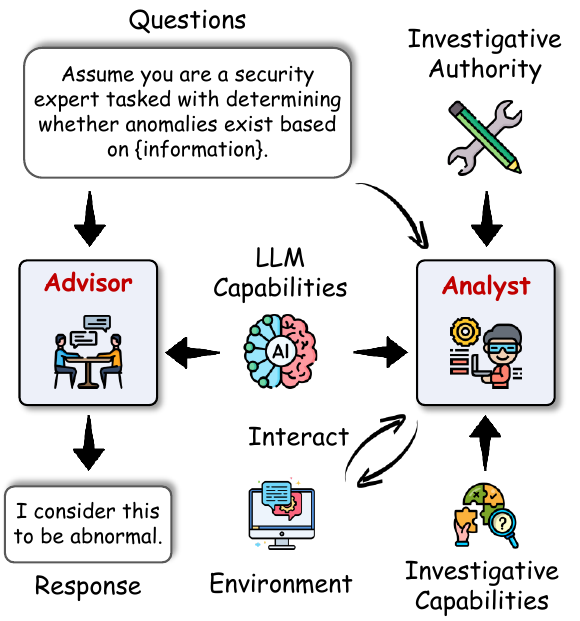}
    \caption{Comparison between the Security Advisor mode and the proposed Active Security Analyst framework.}
    \label{fig:agent-compare}
\end{figure}

As illustrated in Figure~\ref{fig:agent-compare}, We believe that LLMs possessing extensive background knowledge should not merely function as security advisors (judging anomaly probabilities based only on Isolated data). Instead, they should be integrated into the security investigation process as active security analysts, granted actual investigative authority and investigative capabilities. 

\subsection{Threat Model}
We consider a sophisticated adversary that can exploit zero-day vulnerabilities and employ \textit{Living-off-the-Land (LotL)} tactics to evade detection. Such adversaries may mimic benign behavioral patterns to achieve semantic masking. Nevertheless, accomplishing malicious objectives still requires interacting with system resources, which leaves immutable causal dependencies in provenance telemetry. Following prior work~\cite{cheng2024kairos, rehman2024flash}, we treat the OS kernel and its auditing framework as the Trusted Computing Base (TCB) and assume that once generated and collected, audit logs cannot be tampered with, making the provenance data trustworthy throughout our pipeline. We also assume a short benign period to initialize behavioral profiles. Attacks targeting hardware, physical side channels, or the integrity of the TCB are out of scope.

\section{The Entity-Profile-based Detection (EPD)}
\begin{figure*}[t]
    \centering
    \includegraphics[width=0.8\linewidth]{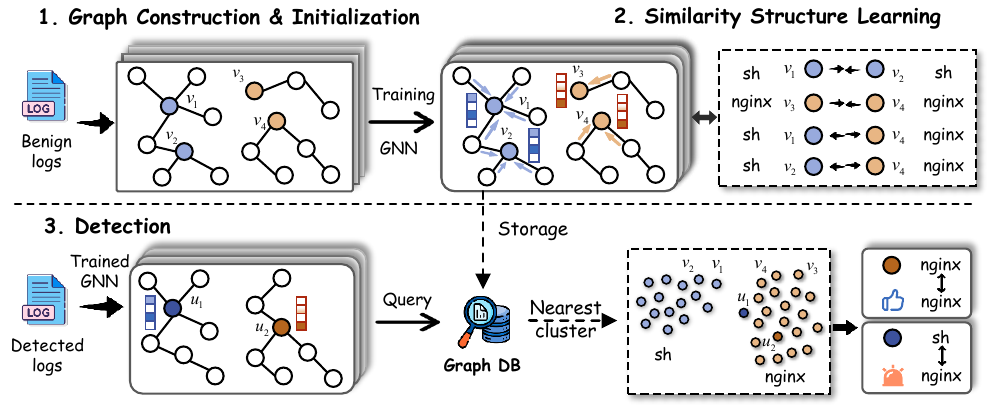}
    \caption{Architecture of the Entity-Profile-based Detection (EPD) module.}
    \label{fig:system_architecture}
\end{figure*}
\section{System Architecture}
ProvAgent comprises two complementary modules that synergize to achieve both breadth and depth in threat investigation. As illustrated in Figure~\ref{fig:system_architecture}, the Entity-Profile-based Detection (EPD) module constructs a cost-effective and efficient P-EDR system that analyzes massive raw audit logs to generate high-quality alerts. As shown in Figure~\ref{fig:agents}, the Multi-Agent Investigation (MAI) module then conducts in-depth analysis by leveraging these alerts as initial leads, performing iterative reasoning with enriched contextual information to identify false positives and reveal missed attacks initially overlooked by the detection system. Ultimately enabling complete attack reconstruction.

The Entity-Profile-based Detection (EPD) module first constructs provenance graphs from massive audit logs, then leverages GNNs to capture behavioral features and generate node embeddings. During training on benign provenance graphs, EPD optimizes embeddings by pulling together nodes with similar identities while pushing apart dissimilar ones. These learned embeddings establish entity profiles that encode identity-behavior patterns of benign entities, serving as baselines for anomaly detection. EPD detects anomalies by assessing identity-behavior consistency, where nodes exhibiting behaviors resembling entities with different identities in the benign repository signal potential threats. Notably, this approach enables explainable attribution, as the matched benign identity provides interpretable explanations for the anomalous behavior. Beyond generating high-quality alerts, EPD maintains a graph database containing node behavioral embeddings and raw provenance data. The alerts serve as investigation leads, while the graph database provides comprehensive forensic materials for subsequent security analysis.

The Multi-Agent Investigation (MAI) module employs a collaborative multi-agent architecture that drives autonomous investigation through a "Hypothesis-Verification" closed-loop framework. This mechanism enables the system to transcend passive alert processing by proactively formulating strategic hypotheses regarding missing attack links or potential false positives. It subsequently directs agents to verify these hypotheses through evidence retrieval to sustain an iterative reasoning process. To ensure rigorous validation, MAI further incorporates a validation mechanism grounded in benign behavioral references. By contrasting suspicious behaviors against a knowledge base of benign profiles, the module can distinguishes operational heterogeneity from genuine attacks. This collaborative and self-correcting approach, supported by a shared investigation repository, mitigates single-point hallucinations and ensures the generation of high-fidelity, analyst-oriented reports.


\subsection{Graph Construction \& Initialization}
ProvAgent ingests raw audit logs from diverse monitoring frameworks such as Linux Audit~\cite{linux_audit_daemon}, Event Tracing for Windows (ETW)~\cite{windows}, CamFlow~\cite{camflow}, and eAudit~\cite{sekar2024eaudit}, and constructs provenance graphs \(\mathcal{G} = (\mathcal{V}, \mathcal{E})\) that capture causal dependencies across system operations. Each node \(v \in \mathcal{V}\) represents a system entity (process, file, socket), while directed edges \(e = (v_i, v_j) \in \mathcal{E}\) encode causal interactions derived from system calls (e.g., \texttt{READ}, \texttt{WRITE}, \texttt{EXECUTE}).

To mitigate the dependency explosion problem~\cite{jiang2025orthrus} caused by high-frequency repetitive operations, we apply edge aggregation to merge redundant interactions. Specifically, multiple edges with identical source, destination, and operation type within a sliding temporal window are collapsed into a single weighted edge, preserving causal semantics while reducing graph complexity.

Node initialization combines three complementary feature encodings to capture multi-faceted behavioral characteristics: semantic attributes, action distributions, and temporal patterns. This multi-modal representation forms the foundation for subsequent discriminative metric learning.

\noindent\textbf{Semantic Encoding.} 
We extract domain-specific semantic attributes from each node based on its entity type, including process name for processes, file paths for files, and IP-port tuples for sockets. To preserve local operational context, we construct structured summaries by concatenating each node's intrinsic attributes with operation types from its 1-hop neighborhood in temporal order. A Word2Vec~\cite{mikolov2013efficient} skip-gram model trained on these summaries learns distributed representations $\mathbf{x}_i^{\text{sem}} \in \mathbb{R}^{d_s}$ that jointly encode semantic attributes and behavioral context.

\noindent\textbf{Action Frequency Encoding.}
We quantify behavioral characteristics through operation frequency distributions. For node $v_i$, we compute a histogram over predefined action categories (e.g., \texttt{READ}, \texttt{WRITE}, \texttt{EXECUTE}). Let $\mathbf{c}_i = [c_1, c_2, \ldots, c_k]$ denote the raw count vector where $c_j$ represents the frequency of action type $j$ associated with $v_i$. To ensure scale invariance and prevent high-volume benign activities from dominating the feature space, we apply $\ell_2$ normalization:

\begin{equation}
    \mathbf{x}_i^{\text{act}} = \frac{\mathbf{c}_i}{\|\mathbf{c}_i\|_2} = \frac{[c_1, c_2, \ldots, c_k]}{\sqrt{\sum_{j=1}^{k} c_j^2}}.
\end{equation}

\noindent\textbf{Temporal Statistics Encoding.}
We characterize temporal regularities through inter-event idle period analysis. For a chronologically ordered event sequence involving node $v_i$ with timestamps $\{t_1, t_2, \ldots, t_m\}$, we derive the idle period sequence $\mathcal{I}_i = \{\Delta t_k = t_{k+1} - t_k \mid k = 1, \ldots, m-1\}$. To summarize the temporal distribution, we extract a statistical vector $\mathbf{s}_i = \left[\min(\mathcal{I}_i), \max(\mathcal{I}_i), \text{mean}(\mathcal{I}_i)\right]$, reflecting activity burstiness and regularity. We apply min-max normalization to project these statistics into the $[0,1]$ range:
\begin{equation}
    \mathbf{x}_i^{\text{tmp}} = \frac{\mathbf{s}_i - \min(\mathbf{s}_i)}{\max(\mathbf{s}_i) - \min(\mathbf{s}_i)}.
\end{equation}

The final initial feature representation $\mathbf{h}_i^{(0)}$ for node $v_i$ is obtained by concatenating the semantic, behavioral, and temporal components:
\begin{equation}
\label{eq:initial_feature}
    \mathbf{h}_i^{(0)} = \mathbf{x}_i^{\text{sem}} \mathbin{\|} \mathbf{x}_i^{\text{act}} \mathbin{\|} \mathbf{x}_i^{\text{tmp}},
\end{equation}
where $\|$ denotes the concatenation operation.

\subsection{Similarity Structure Learning}

We argue that entity identity governs behavioral patterns, such that entities with identical identities should demonstrate consistent or similar behaviors over time. In contrast to previous approaches like ThreaTrace~\cite{wang2022threatrace}, Flash~\cite{rehman2024flash}, and OCR-APT~\cite{aly2025ocr} that employ coarse-grained categories to define identity, we adopt a fine-grained redefinition based on key entity attributes. As shown in Table~\ref{tab:identity}, ProvAgent defines distinct identity criteria for each fundamental entity type in provenance graphs, including subject, file, and netflow.

\begin{table*}[t]
\centering
\renewcommand{\arraystretch}{1.3}
    \begin{tabular}{c|lp{4.1cm}p{6cm}}
    \toprule
    \textbf{Entity Type} & \textbf{Key Attribute} & \textbf{Example} & \textbf{Description} \\
    \midrule
    subject & \makecell{Process name} & \texttt{subject::nginx} & Identical process names indicate the same program logic \\
    \midrule
    netflow & \makecell{Service port} & \texttt{net::80} & Port numbers indicate similar network services or application protocols \\
    \midrule
    file & \makecell{File name} & \texttt{file::nginx.conf} & Path patterns and filenames reflect similar functional roles \\
    \bottomrule
    \end{tabular}
    \caption{Identity criteria for different entity types in provenance graphs. ProvAgent distinguishes entity identities based on fine-grained key attributes rather than coarse-grained categorical labels.}
    \label{tab:identity}
\end{table*}

However, the domain-knowledge-based entity identities result in a large number of distinct identity classes, making conventional node classification tasks inadequate for effectively binding identity with behavioral patterns. To address this challenge, ProvAgent introduces a representation learning framework that integrates graph structural learning with a metric learning objective.

To capture the complex dependencies within provenance graphs, ProvAgent employs a Graph Neural Network (GNN) backbone to extract structural semantic contexts. Specifically, we treat the provenance graph as a directed graph where each system entity acts as a node with initial attributes. The GNN updates node embeddings via an iterative message-passing mechanism. At the $l$-th layer, the representation of node $v_i$ is updated by aggregating information from its local neighborhood:
\begin{equation}
\label{eq:gnn_update}
    \mathbf{h}_i^{(l+1)} = \sigma\left( \mathbf{W}^{(l)} \cdot \text{AGG}\left( \left\{ \mathbf{h}_j^{(l)} \mid v_j \in \mathcal{N}(i) \cup \{v_i\} \right\} \right) \right),
\end{equation}
where $\mathbf{h}_i^{(l)}$ denotes the latent state of node $v_i$ at layer $l$ (with $\mathbf{h}_i^{(0)}$ being the initial feature vector), $\mathcal{N}(i)$ represents the set of immediate neighbors, and $\text{AGG}(\cdot)$ is a permutation-invariant aggregation function (e.g., mean or sum pooling). $\mathbf{W}^{(l)}$ is the learnable weight matrix, and $\sigma(\cdot)$ serves as the non-linear activation function. After $L$ layers of propagation, the final embedding $\mathbf{z}_i = \mathbf{h}_i^{(L)}$ encodes both the intrinsic attributes and the multi-hop structural context of the entity.

To encode the core behavioral patterns associated with specific identities into the node embeddings, ProvAgent employs a contrastive metric learning strategy. The primary objective is to pull together nodes of the same identity while pushing apart nodes of different identities in the latent space. We achieve this with an InfoNCE loss~\cite{ju2024towards}.

For each anchor node $v_i$, we sample a positive node $v_p$ that shares the same identity (or behavioral semantics) and treat the remaining nodes in the mini-batch as negatives. The training objective maximizes the similarity between the anchor and its positive relative to all negatives. Let $\phi_{ij} = s(\mathbf{z}_i, \mathbf{z}_j)/\tau$ denote the temperature-scaled similarity, then:
\begin{equation}
\label{eq:triplet_loss}
    \mathcal{L}_{\text{nce}} = - \sum_{(v_i, v_p) \in \mathcal{P}} \log \frac{\exp(\phi_{ip})}{\exp(\phi_{ip}) + \sum_{v_n \in \mathcal{N}_i} \exp(\phi_{in})},
\end{equation}
where $\mathcal{P}$ denotes the set of anchor-positive pairs, $\mathcal{N}_i$ is the set of negatives for $v_i$ within the mini-batch, $s(\cdot, \cdot)$ is the cosine similarity, and $\tau$ is the temperature parameter. By minimizing this loss, ProvAgent enforces a discriminative embedding space that tightly clusters identical identities while separating distinct ones, thereby binding identity semantics to behavior.

\subsection{Detection}

In contrast to prior approaches, ProvAgent performs anomaly detection grounded in the hypothesis that \emph{entities sharing the same identity exhibit similar behavioral patterns}. It first creates a behavioral reference library by collecting extensive benign audit logs in controlled environments to capture legitimate behaviors across diverse identities. During detection, ProvAgent infers the expected identity of a target node based on its observed behavior and compares it with the node's declared identity attributes. Specifically, any inconsistency where observed actions contradict the declared identity indicates an anomaly. This method facilitates explainable detection by revealing the specific behavioral evidence for each alert. For instance, if a declared \texttt{nginx} process behaves like the \texttt{find} utility, analysts can directly trace the anomaly to the mismatch between the observed file-searching behavior and the declared web server role.

\noindent\textbf{Benign Behavioral Profiling.} To enable zero-shot anomaly detection without prior knowledge of attack signatures, we first establish a comprehensive behavioral baseline by mapping system entities from benign execution traces into the feature space using the pre-trained GNN encoder. 
This process generates a repository of latent embeddings $\mathbf{z}_i \in \mathbb{R}^d$ that encapsulate the legitimate operational patterns of diverse system identities. To support both high-efficiency retrieval and in-depth forensics, we construct a retrieval-augmented knowledge base that indexes these behavioral embeddings alongside their raw provenance metadata.

We formalize the expected behavioral boundary for each semantic identity $c$ (as defined in Table~\ref{tab:identity}) as a high-dimensional hypersphere. Let $\mathcal{Z}_c = \{\mathbf{z}_1^c, \dots, \mathbf{z}_{N_c}^c\}$ denote the set of reference embeddings for a specific identity category in the benign dataset. We characterize the core behavioral pattern of $c$ using a geometric centroid $\boldsymbol{\mu}_c$:
\begin{equation}
\boldsymbol{\mu}_c = \frac{1}{N_c} \sum_{i=1}^{N_c} \mathbf{z}_i^c.
\end{equation}
During the profiling stage, ProvAgent learns identity-specific behavioral deviation patterns from benign data and adopts the corresponding hypersphere radius as an explicit metric to quantify the magnitude of deviation tolerated by each identity class. Accordingly, we define the radius $R_c$ as:
\begin{equation}
R_c = \min \Big\{ r \;\Big|\;
\frac{1}{N_c} \sum_{i=1}^{N_c}
\mathbb{I}\big(\|\mathbf{z}_i^c - \boldsymbol{\mu}_c\|_2 \le r\big)
\ge 1 - \epsilon
\Big\},
\end{equation}
where $\epsilon \in (0,1)$ controls the fraction of benign nodes permitted to fall outside the behavioral boundary. Intuitively, $R_c$ corresponds to the smallest radius that encloses at least a $(1-\epsilon)$ portion of benign samples.

\noindent\textbf{Identity-Consistency Violation Detection.} During online monitoring, ProvAgent transforms threat detection into a geometric containment problem. For an incoming test node $v_j$ with claimed identity $\hat{c}$, we first compute its behavioral embedding $\mathbf{z}_j$ and identify the best-matching identity prototype:
\begin{equation}
c^* = \arg\min_{c} \|\mathbf{z}_j - \boldsymbol{\mu}_c\|_2.
\end{equation}
We detect identity-behavior consistency violations through two complementary criteria: (1) behavioral deviation, where the node's embedding falls outside the benign hypersphere of its claimed identity ($\|\mathbf{z}_j - \boldsymbol{\mu}_{\hat{c}}\|_2 > R_{\hat{c}}$), and (2) identity mismatch, where the best-matching behavioral prototype $c^*$ diverges from the declared identity $\hat{c}$, indicating the entity exhibits behavioral patterns characteristic of a different identity class.

Formally, we define the deviation distance $d_j = \|\mathbf{z}_j - \boldsymbol{\mu}_{\hat{c}}\|_2$ as the Euclidean distance between the node's embedding and its claimed identity prototype. Entities satisfying both $d_j \le R_{\hat{c}}$ and $c^* = \hat{c}$ are classified as benign operational variations. Conversely, violation of either criterion signals behavioral anomalies fundamentally incompatible with the declared identity. 

\section{Multi-Agent Investigation (MAI)}

\begin{figure}[t]
    \centering
    \includegraphics[width=\linewidth]{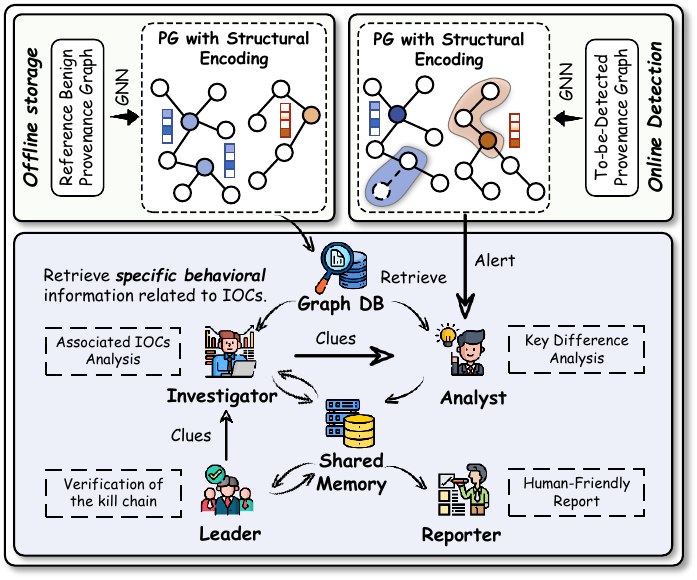}
    \caption{Multi-Agent Investigation (MAI) as a closed-loop provenance-driven investigation pipeline.}
    \label{fig:agents}
\end{figure}

To bridge the gap between automated investigation and expert-level reasoning, we propose a multi-agent investigation framework that elevates raw alerts into comprehensive attack narratives. As illustrated in Figure~\ref{fig:agents}, our architecture orchestrates four specialized agents, namely the \textit{Analyst}, \textit{Investigator}, \textit{Leader}, and \textit{Reporter}, around a centralized investigation repository, which serves as a shared investigation state to keep all agents aligned on validated IOCs, reconstructed attack events, supporting evidence, and strategic hypotheses.

In contrast to workflow-driven systems that follow rigid investigation sequences (e.g., OCR-APT~\cite{aly2025ocr}), MAI adopts a hypothesis-driven closed-loop architecture that coordinates strategic reasoning with evidence-grounded validation through iterative feedback. This design addresses key challenges in autonomous threat investigation along four aspects.

\noindent\textbf{Iterative Hypothesis-Driven Investigation.}
Human analysts typically iterate between hypothesis and verification rather than drawing conclusions in a single pass. By separating strategic reasoning (\textit{Leader}) from tactical validation (\textit{Analyst}, \textit{Investigator}), MAI supports iterative refinement. The \textit{Leader} formulates falsifiable hypotheses about missing attack stages or suspicious IOCs, and the \textit{Analyst} and \textit{Investigator} gather and assess evidence to validate or refute them. This closed-loop process mirrors expert workflows that revise interpretations as new evidence emerges.

\noindent\textbf{Multi-Perspective Analytical Scrutiny.}
MAI implements complementary perspectives through role specialization. The \textit{Investigator} reconstructs atomic operations at the event level, while the \textit{Leader} reasons at the campaign level to identify gaps and inconsistencies. The \textit{Analyst} acts as a validation gatekeeper, preventing both tactical false positives and unsupported strategic speculation from entering the durable IOC set. This separation avoids conflating operational details with campaign-level inference and reduces local bias and globally inconsistent conclusions.

\noindent\textbf{Collaborative Hallucination Mitigation.}
Single-agent LLM pipelines can suffer from error propagation, where early incorrect inferences bias later steps. MAI mitigates this through cross-agent verification. Hypotheses from the \textit{Leader} and leads surfaced by the \textit{Investigator} must pass \textit{Analyst} scrutiny before IOC admission. The \textit{Leader} also performs global consistency checks over accumulated IOCs and events, triggering re-examination when gaps or contradictions arise. A shared investigation repository keeps agents aligned on validated artifacts and supporting evidence, preventing divergent reasoning trajectories.

\noindent\textbf{Evidence-Driven Termination.}
MAI uses evidence-driven termination criteria to avoid premature conclusions and unbounded loops. The investigation stops when either (1) the accumulated IOCs form a coherent kill-chain narrative with no remaining actionable gaps, or (2) all testable hypotheses are exhausted without producing new validated IOCs. By constraining hypotheses with domain rules so that they reference existing IOC metadata and admit falsifiable checks, MAI naturally converges once no further evidence-consistent hypotheses remain.

\subsection{Hypothesis–Verification Orchestration}
MAI operates on typed artifacts and examines and validates IOCs from multiple perspectives, with a centralized investigation repository serving as a shared investigation state. This shared state stores not only validated IOCs but also their supporting provenance evidence and enrichment metadata, and is continuously updated as the investigation evolves. Starting from an alert, MAI proceeds through four stages: \ding{172} MAI validates whether the alert reflects a genuine anomaly; \ding{173} MAI expands from validated IOCs to uncover multi-stage behaviors and surface new leads; \ding{174} MAI synthesizes and corrects the global narrative by pruning spurious IOCs and hypothesizing missing steps; and \ding{175} MAI compiles the final analyst-facing report.

\noindent \textbf{Stage 1: Preliminary IOC identification.}
To narrow the investigation scope, the Analyst first validates the high-confidence alerts generated by the EPD module. For each alerted entity, the Analyst considers both its identity attributes and observed behavioral characteristics, and retrieves corresponding benign behaviors and similar behavior chains from a benign reference repository. By comparing the target entity with benign counterparts, the Analyst determines whether the observed behavioral differences arise from normal operational heterogeneity or from genuine malicious activity, thereby filtering out incidental anomalies caused by legitimate behavior variations. Entities whose behaviors are validated as attack-related are preliminarily identified as IOCs and recorded in the shared repository.

\noindent \textbf{Stage 2: Event-level investigation and controlled expansion.}
Based on the preliminarily identified IOCs, the Investigator conducts event-level analysis over the provenance neighborhood to infer attack events and uncover attack leads, namely attack-related neighboring entities that are not directly revealed by the original alerts. Each lead must be validated by the Analyst before it can be admitted into the persistent IOC set. Once a lead is confirmed as an IOC, the Investigator is re-invoked to infer its associated attack events and to discover new leads. This expansion loop continues until all confirmed IOCs have been fully investigated, thereby improving coverage of multi-step attack activities. All validated findings are ultimately stored in the shared repository to support collaborative analysis across agents.

\noindent \textbf{Stage 3: Strategic reasoning and loop validation.}
Specifically, the Leader performs a global analysis over all preliminarily identified attack behaviors by leveraging the shared repository, and attempts to organize these fragmented behaviors into a coherent attack kill chain. During this process, the Leader reasons about attack steps from an attacker-oriented and strategic perspective, guided by the already inferred attack behaviors. Based on this strategic reasoning and the outcomes of preliminary investigations, the Leader generates explicit hypotheses. Typical hypotheses include that some preliminarily identified attack behaviors or IOCs may be false positives because they cannot be integrated into a plausible kill chain, or that discontinuities in the inferred attack chain indicate missing attack behaviors.

The Leader then feeds these hypotheses back to the Investigator and the Analyst in the form of targeted leads, driving them to conduct focused validation. Hypotheses that are validated through this process are incorporated into the shared repository, while unsupported ones are discarded. To bound the hypothesis space, the Leader uses domain-specific information associated with IOCs to define the scope of hypothesis generation. When all leads are exhausted and the attack chain remains unexplained, the Leader terminates hypothesis generation and treats the remaining attack determinations as false positives. Conversely, when a coherent attack chain can be established, the Leader also stops further hypothesis generation.

\noindent \textbf{Stage 4: Analyst-facing report synthesis.}
Finally, the Reporter transforms the consolidated shared repository into human-readable provenance reports for security analysts, presenting evidence-grounded attack narratives.

\subsection{The Multi-Agents}

\noindent \textbf{Analyst Agent.} 
The Analyst functions as an evidence-based gatekeeper within the investigation pipeline, tasked with validating anomalies. Upon encountering an alert or suspicious lead, the Analyst conducts a fine-grained behavioral comparison by retrieving similar benign entities from a reference repository using a dual-mode matching strategy. This approach combines attribute-based string matching to locate entities with comparable identity characteristics and embedding-based vector similarity search to identify behaviorally analogous counterparts. By contrasting the provenance traces of the target against these benign baselines, the Analyst discriminates genuine malicious deviations from benign operational heterogeneity. Consequently, only entities exhibiting behavioral anomalies that cannot be reconciled with legitimate variations are escalated to the investigation repository. This rigorous filtering process eliminates spurious alerts arising from normal system diversity and establishes a high-confidence set of threat indicators grounded in evidence from the benign knowledge base.

\noindent \textbf{Investigator Agent.} 
The Investigator conducts event-level forensic analysis to uncover the tactical execution details of confirmed IOCs. For each validated IOC, the Investigator traverses its temporal event sequences and provenance neighborhood to reconstruct atomic attack operations, such as file exfiltration, privilege escalation, or lateral movement attempts. Beyond documenting attack mechanics, the Investigator identifies potentially related neighboring entities (e.g., spawned processes, accessed files, contacted sockets) as new investigative leads. These leads are routed back to the Analyst for validation, establishing a controlled expansion loop that progressively uncovers multi-stage attack activities while preventing unbounded false-positive propagation. The Investigator also annotates each IOC with kill chain stage mappings and detailed event descriptions, enriching the contextual understanding for downstream strategic reasoning.

\noindent \textbf{Leader Agent.} 
The Leader performs strategic-level coordination and global coherence validation. Operating on the collective set of validated IOCs and their enriched metadata, the Leader attempts to synthesize a complete attack narrative by mapping fragmented behaviors onto a structured kill chain framework (e.g., reconnaissance, initial access, execution, persistence, lateral movement, exfiltration). When gaps or inconsistencies emerge in the inferred attack sequence, the Leader generates strategic hypotheses from an adversarial perspective: missing intermediate steps that should logically exist (e.g., if lateral movement is detected but no credential harvesting is observed, the Leader hypothesizes potential credential theft activities), or potentially spurious IOCs that cannot be coherently integrated into the narrative (indicating possible false positives). These hypotheses are formulated as directed investigative tasks and dispatched to the Investigator and Analyst for targeted verification, forming a closed-loop refinement cycle. The Leader terminates hypothesis generation when either a coherent end-to-end attack chain is established or all plausible hypotheses have been exhausted, thereby preventing infinite investigative loops.

\noindent \textbf{Reporter Agent.} 
The Reporter transforms the investigation repository into analyst-ready reports by organizing validated IOCs into a coherent attack narrative, constructing a provenance graph grounded in event evidence, and summarizing actionable remediation guidance.

\begin{table}[h]
\centering
\small
\setlength{\tabcolsep}{5pt}
\resizebox{\linewidth}{!}{%
    \begin{tabular}{l l r r r r}
    \toprule
    \textbf{Dataset} & \textbf{System} & \textbf{TP} & \textbf{FP} & \textbf{TN} & \textbf{FN} \\
    \midrule

    \multirow{8}{*}{\begin{tabular}{c}
    DARPA E3 \\
    CADETS
    \end{tabular}}
    & ProvAgent   & 19 & 1,941 & 266,144 & 53 \\
    & Threatrace  & 61  & 252,117    & 15,968      & 7  \\
    & Kairos      & 0  & 9    & 268,076      & 68  \\
    & Flash       & 13  & 2,381    & 265,704      & 55  \\
    & MAGIC       & 63  & 79,766    & 188,319      & 5  \\
    & Orthrus     & 10  & 0    & 268,085      & 58  \\
    & OCR-APT     & 53  & 83,779    & 184,306      & 15  \\
    \midrule

    \multirow{8}{*}{\begin{tabular}{c}
    DARPA E3 \\
    THEIA
    \end{tabular}}
    & ProvAgent   & 71  & 6,574    & 692,603      & 47  \\
    & Threatrace  & 88  & 671,883    & 27,294      & 30  \\
    & Kairos      & 4  & 0    & 699,177      & 114  \\
    & Flash       & 22  & 32,082    & 667,095      & 96  \\
    & MAGIC       & 115  & 394,906    & 304,271      & 3  \\
    & Orthrus     & 8  & 0    & 699,177      & 110  \\
    & OCR-APT     & 39  & 121,339    & 577,838      & 79  \\
    \midrule

    \multirow{8}{*}{\begin{tabular}{c}
    DARPA E5 \\
    CADETS
    \end{tabular}}
    & ProvAgent   & 39 & 10,272 & 3,100,983 & 84 \\
    & Threatrace  & 91  & 3,104,018    & 7,237      & 32  \\
    & Kairos      & 0  & 6    & 3,111,249      & 123  \\
    & Flash       & 45  & 33,941    & 3,077,314      & 78  \\
    & MAGIC       & 123  & 3,110,714    & 541      & 0  \\
    & Orthrus     & 1  & 5    & 3,111,250      & 122  \\
    & OCR-APT     & 101  & 3,109,967    & 1,288      & 22  \\
    \midrule

    \multirow{8}{*}{\begin{tabular}{c}
    DARPA E5 \\
    THEIA
    \end{tabular}}
    & ProvAgent   & 20  & 3,131    & 744,252      & 49  \\
    & Threatrace  & 66  & 739,322    & 8,061      & 3  \\
    & Kairos      & 0  & 2    & 747,381      & 69  \\
    & Flash       & 43  & 295,729    & 451,654      & 26  \\
    & MAGIC       & 1  & 296,554    & 450,829      & 68  \\
    & Orthrus     & 2  & 0    & 747,383      & 67  \\
    & OCR-APT     & 57  & 127,948    & 619,435      & 12  \\
    \bottomrule
    \end{tabular}
}
\caption{Detection performance comparison between the EPD module of ProvAgent and sota methods on E3 and E5 datasets.}
\label{tab:darpa_e_detection}
\end{table}
\section{Evaluation}
In this section, experiments are performed to validate ProvAgent's advantage and answer the following key research questions:

\begin{itemize}[leftmargin=*,topsep=0pt,itemsep=0pt,partopsep=0pt,parsep=0pt]
  \item \textbf{RQ1:} How effective is ProvAgent at anomaly detection?
  \item \textbf{RQ2:} How do hyperparameters affect detection performance?
  \item \textbf{RQ3:} How effective is ProvAgent at attack investigation and reconstruction?
  \item \textbf{RQ4:} How do different components of ProvAgent contribute to its detection performance?
  \item \textbf{RQ5:} What is the operational overhead of ProvAgent under realistic workloads?
  \item \textbf{RQ6:} How robust is ProvAgent against mimicry attacks?
\end{itemize}

\subsection{Experimental Setups}
\noindent \textbf{Datasets.} We evaluate on three widely used enterprise-scale provenance benchmarks, DARPA Transparent Computing Engagement 3 (E3), Engagement 5 (E5)~\footnote{The CLEARSCOPE dataset in the DARPA corpus collects logs from the Android framework. We compared the labels provided by Orthrus with the official DARPA documentation and identified multiple inconsistencies. We therefore exclude the CLEARSCOPE dataset from our evaluation.}, and Operationally Transparent Cyber (OPTC). E3 and E5 were collected during DARPA Transparent Computing red team engagement exercises and record multi-host, whole-system activity in which multi-stage APT campaigns are embedded within substantial benign background workloads. E5 is generally more challenging than E3 due to its larger scale and more complex, stealthier attack execution. OPTC further increases operational realism and data volume by capturing long-running enterprise activity with severe class imbalance, while red team malicious behaviors are injected into continuous benign telemetry. For evaluation, we adopt the same labeling schemes as Orthrus.

\noindent \textbf{Baseline Methods.}
We compare against seven state-of-the-art provenance based threat detection systems, including Kairos~\cite{cheng2024kairos}, ThreaTrace~\cite{wang2022threatrace}, Flash~\cite{rehman2024flash}, MAGIC~\cite{jia2024magic}, Orthrus~\cite{jiang2025orthrus} and OCR-APT~\cite{aly2025ocr}.
We exclude methods that do not support node-level detection, such as Unicorn and StreamSpot, because they cannot be evaluated at the same granularity as our approach.
We exclude supervised methods that require labeled attack samples (e.g., Slot and ATLAS) because the datasets are extremely imbalanced, with only a small fraction of attack-related nodes, making supervised learning unreliable. We also omit heuristic systems driven mainly by threat intelligence or expert rules (e.g., Holmes and Connan), since their manually crafted knowledge hinders fair and reproducible comparison.

\noindent \textbf{Implementation.}
All training and evaluation were conducted on a server running Ubuntu 24.04, equipped with a dual Intel Xeon Platinum 8558 CPU, 1024 GB of memory, and an NVIDIA H200 GPU with 141GB of memory.

\subsection{Attack Detection Performance (RQ1)}

\begin{table}[t]
\centering
\small
\setlength{\tabcolsep}{5pt}
\resizebox{\linewidth}{!}{%
    \begin{tabular}{l l r r r r}
    \toprule
    \textbf{Dataset} & \textbf{System} & \textbf{TP} & \textbf{FP} & \textbf{TN} & \textbf{FN} \\
    \midrule

    \multirow{8}{*}{\begin{tabular}{c}
    OPTC \\
    H051
    \end{tabular}}
    & ProvAgent   & 49 & 8,902 & 1,461,608 & 65 \\
    & Threatrace  & 91  & 1,392,711    & 77,799      & 23  \\
    & Kairos      & 1  & 2    & 1,470,508      & 113  \\
    & Flash       & 12  & 10,031    & 1,460,479      & 102  \\
    & MAGIC       & 90  & 241,724    & 1,228,786      & 24  \\
    & Orthrus     & 0  & 0    & 1,470,506      & 114  \\
    & OCR-APT     & 56  & 788,239    & 682,271      & 58  \\
    \midrule

    \multirow{8}{*}{\begin{tabular}{c}
    OPTC \\
    H201
    \end{tabular}}
    & ProvAgent   & 102  & 6,988    & 1,428,041      & 2,803  \\
    & Threatrace  & 2,905  & 1,435,029    & 0      & 0  \\
    & Kairos      & 1  & 0    & 1,435,029      & 2,904  \\
    & Flash       & 1,473  & 240,117    & 1,194,912      & 1,432  \\
    & MAGIC       & 1,013  & 1,222,313    & 212,716      & 1,892  \\
    & Orthrus     & 1  & 4    & 1,435,025      & 2,904  \\
    & OCR-APT     & 1,373  & 79,144    & 1,355,885      & 1,532  \\
    \midrule

    \multirow{8}{*}{\begin{tabular}{c}
    OPTC \\
    H501
    \end{tabular}}
    & ProvAgent   & 89  & 9,876    & 1,425,153   & 660  \\
    & Threatrace  & 2  & 42   & 1,434,987      & 747  \\
    & Kairos      & 1  & 2    & 1,435,027      & 748  \\
    & Flash       & 221  & 215,131    & 1,219,898      & 528  \\
    & MAGIC       & 177  & 253,984    & 1,181,045      & 572  \\
    & Orthrus     & 1  & 4    & 1,435,025      & 748  \\
    & OCR-APT     & 89  & 9,876    & 1,425,153      & 660  \\
    \bottomrule
    \end{tabular}
}
\caption{Detection performance comparison between the EPD module of ProvAgent and sota methods on OPTC datasets.}
\label{tab:darpa_optc_detection}
\end{table}
\begin{figure*}[h]
    \centering
    \includegraphics[width=\linewidth]{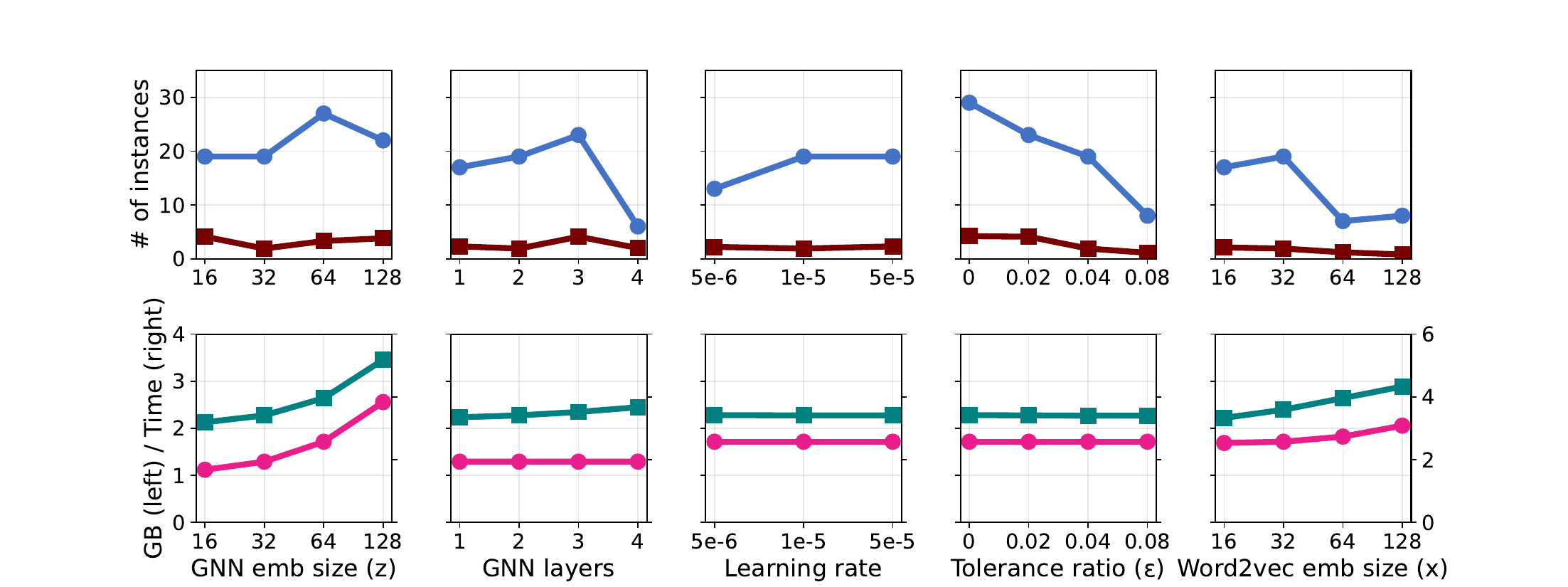}
    \caption{Hyperparameter study of ProvAgent on CADETS-E3. We show \textcolor[HTML]{4472C4}{\rule{0.8em}{0.8em}} TP, \textcolor[HTML]{780000}{\rule{0.8em}{0.8em}} FP (k), \textcolor[HTML]{E91E8C}{\rule{0.8em}{0.8em}} memory consumption (GB), and \textcolor[HTML]{008080}{\rule{0.8em}{0.8em}} inference time (s).}
    \label{fig:hyperparameter}
\end{figure*}

In the first stage, ProvAgent leverages the EPD module to perform anomaly detection and generate high-quality alerts. As shown in Tables~\ref{tab:darpa_e_detection} and~\ref{tab:darpa_optc_detection}, the overly strict detection strategies adopted by Orthrus lead to extremely low detection rates for real attacks, and in some scenarios even result in zero true positives. Consequently, these systems fail to provide any actionable investigative leads for security analysts. In contrast, while ThreatTrace, Flash, MAGIC, and OCR-APT are able to detect certain anomalous behaviors, they also produce an overwhelming number of false positives, imposing a substantial burden on downstream security analysis. By comparison, ProvAgent successfully captures critical attack-related information with a significantly lower false-positive cost. This advantage stems from abandoning conventional edge-reconstruction and node-reconstruction paradigms in favor of a fine-grained identity–behavior binding framework, which performs anomaly detection through identity profiling and thereby mitigates misinterpretations of identity semantics and stereotypical assumptions about behavior.

We further analyze the false positives generated by ProvAgent and observe that most of them arise from behaviorally similar processes being misidentified due to shared low-level system calls. For example, the \texttt{mv} and \texttt{sleep} processes exhibit nearly identical invocation patterns at the syscall level, which can lead to identity confusion under behavior-based recognition. A more detailed analysis of these cases is provided in the appendix~\ref{appendix:similar}.

\subsection{Hyperparameters Study (RQ2)}

We conduct a systematic ablation study to examine how key hyperparameters affect the detection effectiveness and operational cost of \textsc{ProvAgent}. We vary one hyperparameter at a time while holding all others fixed, and measure its impact on true positives (TP), false positives (FP), memory footprint, and inference latency. The results are summarized in Figure~\ref{fig:hyperparameter}.

\noindent \textbf{GNN Embedding Dimension ($z$).}
The embedding dimension determines the representational capacity of the graph neural network and its computational overhead. We evaluate $z \in \{16, 32, 64, 128\}$ and find that $z=32$ achieves the best balance between detection accuracy and resource consumption. Smaller dimensions fail to capture complex multi-stage attack patterns, while larger dimensions introduce unnecessary overhead and show early signs of overfitting without accuracy gains.

\noindent \textbf{Number of GNN Layers.}
The number of GNN layers controls the receptive field over the provenance graph. Shallow models fail to capture long-range dependencies across causal chains, while increasing depth beyond a moderate level leads to over-smoothing and optimization instability. A two-layer configuration achieves the most robust performance.

\noindent \textbf{Learning Rate.}
We find that $1\times10^{-5}$ achieves the best trade-off between convergence speed and detection performance. In contrast, smaller learning rates converge slowly, while larger ones cause loss oscillation and degrade generalization.

\noindent \textbf{Tolerance Ratio ($\varepsilon$).}
The tolerance ratio $\varepsilon$ controls the system’s sensitivity to deviations from benign behavior. Across $\varepsilon \in {0, 0.02, 0.04, 0.08}$, we find that $\varepsilon = 0.02$ yields the best overall performance. Smaller tolerance values lead to increased false positives, as a more permissive policy is adopted when constructing identity profiles. In contrast, overly large tolerance values may cause subtle attack behaviors to be missed.

\noindent \textbf{Word2Vec Embedding Dimension ($x$).}
The Word2Vec embedding dimension determines the semantic expressiveness of node attributes, such as process names and file paths. Across $x \in {16, 32, 64, 128}$, we observe that a dimension of 32 achieves the best detection performance. Increasing the embedding dimension beyond this point does not yield further improvements. This is because ProvAgent performs anomaly detection under an identity–behavior binding framework: overly rich initial embeddings may allow the model to directly infer entity identities from attribute information, reducing its reliance on behavioral patterns. Conversely, when the embedding dimension is too small, insufficient attribute information limits the model’s ability to distinguish behavior modules that differ only in attributes, leading to less accurate behavior-based identity recognition. We provide a more detailed analysis of this phenomenon in the appendix~\ref{appendix:similar}.

These results indicate that ProvAgent is robust to hyperparameter variations and remains effective at identifying attacks across a wide range of configurations.

\begin{table*}[h]
\centering
\small
\setlength{\tabcolsep}{6.5pt}
\begin{tabular}{l l l l l c c}
\toprule
\textbf{Base Model} & \textbf{Dataset} & \textbf{System} & \textbf{Attack Date} & \textbf{Description} & \textbf{IOCs} & \textbf{Detected APT Stages} \\
\midrule

\multirow{8}{*}{DeepSeek-V3.2}
& \multirow{4}{*}{\shortstack{CADETS \\ E3}}
& \multirow{3}{*}{\textbf{ProvAgent (ours)}} & 2018-04-06 & \multirow{3}{*}{Nginx Backdoor} & 2/4/8 & \textcolor{n}{IC}, \textcolor{y}{C\&C}, \textcolor{y}{PE}, \textcolor{y}{MP} \\
&                              &             & 2018-04-12 &                                             & 3/7/7 & \textcolor{y}{IR}, \textcolor{y}{C\&C}, \textcolor{y}{PE}, \textcolor{y}{LM}, \textcolor{y}{MP} \\
&                              &             & 2018-04-13 &                                             & 2/5/7 & \textcolor{n}{IR}, \textcolor{y}{C\&C}, \textcolor{y}{PE}, \textcolor{y}{MP} \\
\cmidrule(lr){3-7}
&                              & OCR-APT~\cite{aly2025ocr}     & 2018-04-12 & Nginx Backdoor & 4/1/7 & \textcolor{n}{IR}, \textcolor{n}{C\&C}, \textcolor{n}{PE}, \textcolor{n}{LM}, \textcolor{y}{MP} \\
\cmidrule(lr){2-7}

& \multirow{4}{*}{\shortstack{THEIA \\ E3}}
& \multirow{2}{*}{\textbf{ProvAgent (ours)}} & 2018-04-10 & Firefox Backdoor & 3/8/10 & \textcolor{y}{IC}, \textcolor{y}{C\&C}, \textcolor{y}{PE}, \textcolor{y}{MP} \\
&                              &             & 2018-04-12 & Browser extension & 4/9/17 & \textcolor{y}{IC}, \textcolor{y}{IR}, \textcolor{y}{C\&C}, \textcolor{y}{PE}, \textcolor{y}{MP}, \textcolor{y}{CT} \\
\cmidrule(lr){3-7}
&                              & OCR-APT~\cite{aly2025ocr}     & 2018-04-12 & Browser extension & 9/3/17 & \textcolor{n}{IC}, \textcolor{y}{IR}, \textcolor{n}{C\&C}, \textcolor{n}{PE}, \textcolor{y}{MP}, \textcolor{n}{CT} \\

\midrule

\multirow{8}{*}{gpt-5.2}
& \multirow{4}{*}{\shortstack{CADETS \\ E3}}
& \multirow{3}{*}{\textbf{ProvAgent (ours)}} & 2018-04-06 & \multirow{3}{*}{Nginx Backdoor} & 2/5/8 & \textcolor{n}{IC}, \textcolor{y}{C\&C}, \textcolor{y}{PE}, \textcolor{y}{MP} \\
&                              &             & 2018-04-12 &                                             & 3/7/7 & \textcolor{y}{IR}, \textcolor{y}{C\&C}, \textcolor{y}{PE}, \textcolor{y}{LM}, \textcolor{y}{MP} \\
&                              &             & 2018-04-13 &                                             & 2/4/7 & \textcolor{n}{IR}, \textcolor{y}{C\&C}, \textcolor{y}{PE}, \textcolor{y}{MP} \\
\cmidrule(lr){3-7}
&                              & OCR-APT~\cite{aly2025ocr}     & 2018-04-12 & Nginx Backdoor & 4/2/7 & \textcolor{y}{IR}, \textcolor{n}{C\&C}, \textcolor{n}{PE}, \textcolor{n}{LM}, \textcolor{y}{MP} \\
\cmidrule(lr){2-7}

& \multirow{4}{*}{\shortstack{THEIA \\ E3}}
& \multirow{2}{*}{\textbf{ProvAgent (ours)}} & 2018-04-10 & Firefox Backdoor & 3/8/10 & \textcolor{y}{IC}, \textcolor{y}{C\&C}, \textcolor{y}{PE}, \textcolor{y}{MP} \\
&                              &             & 2018-04-12 & Browser extension & 4/16/17 & \textcolor{y}{IC}, \textcolor{y}{IR}, \textcolor{y}{C\&C}, \textcolor{y}{PE}, \textcolor{y}{MP}, \textcolor{y}{CT} \\
\cmidrule(lr){3-7}
&                              & OCR-APT~\cite{aly2025ocr}     & 2018-04-12 & Browser extension & 9/4/17 & \textcolor{y}{IC}, \textcolor{y}{IR}, \textcolor{n}{C\&C}, \textcolor{n}{PE}, \textcolor{y}{MP}, \textcolor{n}{CT} \\

\bottomrule
\end{tabular}
\caption{Comparison of IoC and APT stage investigation performance across different base LLMs (including DeepSeek-V3.2 and gpt-5.2) on DARPA E3. The table shows the number of detected IOCs and APT attack stages, with total counts in parentheses. Detected stages are highlighted in green, while missed stages are shown in red. In the IOCs column, the three values denote the number of detected IOCs from initial detection, the number of detected IOCs after investigation, and the ground-truth IOC count, respectively.}
\label{tab:base_model_comparison}
\end{table*}

\subsection{Attack Investigation Performance (RQ3)}

To ensure a fair comparison, we re-executed OCR-APT~\footnote{OCR-APT only verifies a subset of attacks in each dataset. Although we attempted to reproduce it on other scenarios, the reports did not reveal any genuine attacks. We attribute this to our lack of expertise and thus omit these results from Table~\ref{tab:base_model_comparison}.} using its publicly available code~\cite{ocr_apt_code}, replacing only the underlying base LLM models to match our evaluation setup. We then compare the attack reports generated by ProvAgent and OCR-APT against the ground-truth reports provided by DARPA~\cite{darpa2018tc3_groundtruth}. To enable a fairer comparison, we treat duplicated information as a single IOC, as illustrated by the port-scanning activity performed by the \texttt{test} process on the CADETS-E3 dataset shown in Figure~\ref{fig:cadets6}.

We evaluate the investigation capability on the CADETS E3 and THEIA E3 datasets using DeepSeek-V3.2 and GPT-5.2 as the base LLMs. Our evaluation examines investigation performance with respect to detected indicators of compromise (IOCs) and detected advanced persistent threat (APT) stages. The APT stages include Initial Compromise (IC), Internal Reconnaissance (IR), Command and Control (C\&C), Privilege Escalation (PE), Lateral Movement (LM), Maintain Persistence (MP), Data Exfiltration (DE), and Covering Tracks (CT). 
Across multiple attacks in both datasets, ProvAgent is able to recover an almost complete narrative of each attack and cover the majority of APT stages (highlighted in green). 
Moreover, ProvAgent produces concise, analyst-oriented reports~\footnote{Full versions of the reports are available at \url{https://github.com/Win7ery/ProvAgent}}, including compact attack graphs provided in Appendix~\ref{appendix:attack_reconstruction}, which are automatically generated by ProvAgent without any human intervention.

As shown in Table~\ref{tab:base_model_comparison}, OCR-APT exhibits a declining IOC count after investigation. This behavior is largely due to its lenient detection policy combined with a static, purely reductive filtering mechanism. Without proactive exploration, OCR-APT mainly prunes the initial alerts, and therefore the final IOC count remains strictly below its detection output. In contrast, ProvAgent provides a dual benefit. It not only removes false positives to streamline the report, but also treats validated alerts as anchors for expanded investigation and uncovers substantial missed detections. Starting from the initially flagged anomalous nodes, ProvAgent increases the number of detected IOCs by 160.7\% after investigation, whereas OCR-APT achieves only -61.5\%.

\begin{table}[h]
\centering
\small
\setlength{\tabcolsep}{7pt}
\begin{tabular}{llcc}
\toprule
\textbf{Attack Date} &  & \textbf{IOCs} & \textbf{Detected APT Stages} \\
\midrule
\multirow{2}{*}{2018--04--06}
 & w/o SBR & 8(5) & \textcolor{n}{IC}, \textcolor{y}{C\&C}, \textcolor{y}{PE}, \textcolor{y}{MP} \\
 & w/o SIR & 8(4) & \textcolor{n}{IC}, \textcolor{y}{C\&C}, \textcolor{y}{PE}, \textcolor{y}{MP} \\
\midrule
\multirow{2}{*}{2018--04--12}
 & w/o SBR & 7(7) & \textcolor{y}{IR}, \textcolor{y}{C\&C}, \textcolor{y}{PE}, \textcolor{y}{LM}, \textcolor{y}{MP} \\
 & w/o SIR & 7(6) & \textcolor{y}{IR}, \textcolor{y}{C\&C}, \textcolor{y}{PE}, \textcolor{y}{LM}, \textcolor{y}{MP} \\
\midrule
\multirow{2}{*}{2018--04--13}
 & w/o SBR & 7(6) & \textcolor{n}{IR}, \textcolor{y}{C\&C}, \textcolor{y}{PE}, \textcolor{y}{MP} \\
 & w/o SIR & 7(5) & \textcolor{n}{IR}, \textcolor{y}{C\&C}, \textcolor{y}{PE}, \textcolor{y}{MP} \\
\bottomrule
\end{tabular}
\caption{Ablation results on CADETS-E3 when disabling similar-behavior retrieval (w/o SBR) or same-identity retrieval (w/o SIR) in MAI, using DeepSeek-V3.2 as the base model.}
\label{tab:ablation}
\end{table}

\subsection{Ablation Study (RQ4)}


In ProvAgent, the EPD module serves as the retrieval backend for the MAI module, supporting both same-identity retrieval and similar-behavior retrieval.
To validate the effectiveness of the collaboration between the two core modules in ProvAgent, we conduct an ablation study by selectively removing the \emph{similar-behavior retrieval} and the \emph{same-identity retrieval} components from the MAI module. Using DeepSeek-V3.2 as the base LLM, we evaluate the resulting variants on CADETS-E3, as summarized in Table~\ref{tab:ablation}.

We observe that removing either similar-behavior retrieval or same-identity retrieval negatively impacts the MAI investigation process. Specifically, similar-behavior retrieval facilitates the identification of legitimate behavioral heterogeneity by providing broad cross-entity references, while same-identity retrieval establishes entity-specific baselines that prevent the Analyst from overlooking identity-specific malicious deviations due to an over-reliance on generic benign patterns.

To further assess the critical role of these retrieval mechanisms, we simultaneously removed both components from the Analyst. Under this setting, MAI fails to converge during the alert-driven investigation, entering a degenerate expansion loop that eventually exhausts computational resources. An analysis of the question–answer traces reveals that, in the absence of external knowledge, the Analyst tends to overinterpret alerts as anomalous, given that real-world audit logs inherently contain irregular operations. Consequently, the Analyst repeatedly validates abnormal entities surfaced by the Investigator as true attacks, thereby continuously driving further expansion. This behavior causes the investigation to stall in the second stage, where recursive expansion dominates, effectively impeding the Leader from performing necessary interventions and corrections.

\begin{table}[t]
  \centering
  \small
  \setlength{\tabcolsep}{5pt}
  \resizebox{\linewidth}{!}{
  \begin{tabular}{l l l l l l}
    \toprule
    \textbf{Dataset} & \textbf{Model} & {\textbf{Input/Output}} & \textbf{Price} & \textbf{Time} \\
    \midrule

    \multirow{2}{*}{CADETS-E3}
      & DeepSeek-V3.2 & 0.662M/0.113M & \$0.24 & 0.73~h \\
      & gpt-5.2       & 1.122M/0.278M & \$5.85 & 1.46~h \\
    \midrule

    \multirow{2}{*}{THEIA-E3}
      & DeepSeek-V3.2 & 1.761M/0.228M  & \$1.21 & 0.6~h \\
      & gpt-5.2       & 2.128M/0.404M  & \$9.38 & 0.78~h \\
    \bottomrule
  \end{tabular}
    }
  \caption{Token usage, cost, and runtime across datasets and models.}
  \label{tab:cost_time}
\end{table}
\subsection{System Overhead (RQ5)}
We analyze the overhead of ProvAgent from two complementary dimensions: the system resource consumption of EPD module and the cost of MAI module.

\noindent \textbf{EPD module Overhead.} The EPD module in ProvAgent adopts a GNN–based detection scheme. As a result, the efficiency overhead during detection primarily depends on the depth of the GNN and the embedding dimensionality. As illustrated in Figure~\ref{fig:hyperparameter}, we evaluate the runtime efficiency and memory footprint of the EPD module under varying model configurations.

\noindent \textbf{MAI module Overhead.} We evaluate the overhead of MAI using four days of logs from CADETS-E3 and three days of logs from THEIA-E3, with results summarized in Table~\ref{tab:cost_time}. Across both datasets, ProvAgent completes in-depth investigations at a minimum cost of \$0.06 per day and finishes the daily reasoning workload within an average of 0.252 hours. This efficiency is enabled by our design, which bounds investigation growth by limiting evidence retrieval and enforcing per-lead re-validation. As a result, uncontrolled hypothesis chaining is avoided and token usage remains linear in the number of validated IOCs.

\subsection{Robustness Against Mimicry Attacks (RQ6)}
\begin{figure}
    \centering
    \includegraphics[width=\linewidth]{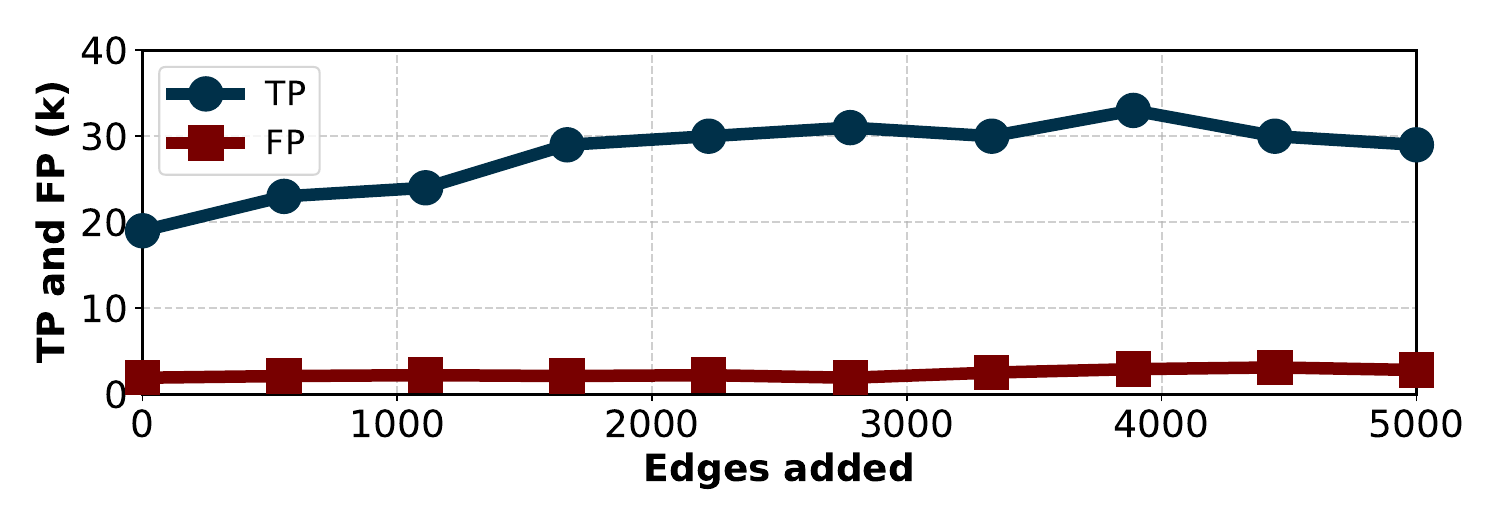}
    \caption{Resilience against adversarial attacks.}
    \label{fig:adversary}
\end{figure}

For provenance-graph-based APT detection systems, mimicry attacks typically evade detection by tampering with raw audit data or injecting benign structures, thereby making malicious activity resemble normal behavior. To evaluate the robustness of ProvAgent under adversarial settings, we adopt the mimicry attack techniques proposed by Goyal \emph{et al.}~\cite{goyal2023sometimes}. These attacks insert benign substructures into attack graphs to deliberately distort local neighborhood distributions, creating deceptive similarity between malicious nodes and benign provenance graphs. Prior work has shown that several PIDS, including StreamSpot, Unicorn, and ProvDetector, are highly vulnerable under such attacks.

We apply Goyal \emph{et al.}~\cite{goyal2023sometimes} publicly released implementation to manipulate the E3-CADETS dataset and evaluate the impact of mimicry attacks on ProvAgent’s detection performance. The results, shown in Figure~\ref{fig:adversary}, indicate that ProvAgent maintains stable detection performance under adversarial manipulation, demonstrating strong robustness.

This robustness stems from identity–behavior binding. Unlike reconstruction-based detectors that judge anomalies via latent-space similarity, ProvAgent learns identity-specific behavioral semantics and permissible variation ranges through contrastive learning. Consequently, benign injections that introduce identity-inconsistent behaviors amplify rather than suppress detection confidence.

\section{Discussion and Limitations}

\noindent\textbf{Benign Baseline Dependency.}
ProvAgent assumes a clean benign period to learn identity-specific behavioral profiles, as is common in anomaly-based systems. This can be challenging when historical telemetry is limited or compromise is suspected. Mitigations include generating benign traces in controlled sandboxes and using transfer learning to reuse profiles across similar environments. Future work could explore federated learning to build shared behavioral baselines across trusted deployments while preserving privacy.

\noindent\textbf{Campaign Segmentation Across Time Windows.} 
ProvAgent currently assumes that each day’s alerts belong to a single campaign and are investigated within an independent daily window. While this simplifies reasoning, real-world APT activity often violates this assumption. Multiple unrelated campaigns may occur within the same day, causing the \textit{Leader} to conflate distinct IOCs into an incorrect narrative. Conversely, a single long-running campaign may span multiple days, leading to fragmented kill-chain reconstruction when days are analyzed in isolation. Future work could incorporate cross-day correlation to link related IOCs across windows and apply clustering or attribution methods to separate concurrent campaigns within the same period.

\noindent\textbf{Kill Chain Rigidity.}
The \textit{Leader} maps behaviors to predefined kill-chain taxonomies (e.g., MITRE ATT\&CK), which improves interpretability and narrative synthesis but can be brittle for novel attacks that deviate from established stage progressions. This limitation is more pronounced for emerging techniques or campaigns that reorder or blur stages (e.g., persistence-first intrusions). 

\section{Related Work}

\noindent \textbf{Anomaly Detection.}
Provenance-based intrusion detection has evolved through three major paradigms. 
\textbf{Heuristic-based methods}, such as Sleuth~\cite{hossain2017sleuth} and Holmes~\cite{milajerdi2019holmes}, construct attack patterns using predefined expert rules or by mapping system activities to known TTPs in knowledge bases such as MITRE ATT\&CK~\cite{mitre_attack}. 
Tactical~\cite{hassan2020tactical} and CONAN~\cite{xiong2020conan} further encode domain knowledge into handcrafted scoring rules to identify suspicious behaviors. 
CAPTAIN~\cite{wang2025incorporating} extends this line of work by incorporating gradient-based optimization to automatically derive adaptive matching rules. 
\textbf{Anomaly-based methods} detect intrusions by modeling benign behavioral baselines from provenance graphs and flagging deviations from normal patterns. 
ThreatTrace~\cite{wang2022threatrace} learns node-level representations with heterogeneous GraphSAGE~\cite{hamilton2017inductive}. Flash~\cite{rehman2024flash} utilizes a GNN model with positional encoding~\cite{vaswani2017attention} to detect anomalous nodes based on type prediction. Kairos~\cite{cheng2024kairos} leverages temporal GNNs~\cite{rossi2020temporal} to capture evolving behaviors. SLOT~\cite{qiao2025slot} further formulates provenance analysis as a graph reinforcement learning problem~\cite{darvariu2024graph}.
\textbf{LLM-based methods} seek to exploit the cross-domain reasoning capabilities of large language models for threat detection. 
PROVSEEK~\cite{mukherjee2025llm} adopts a multi-agent design that combines LLM-driven investigation with external threat intelligence to analyze raw audit logs.
OMNISEC~\cite{cheng2025omnisec} uses retrieval-augmented generation (RAG) over external intelligence and benign knowledge bases to improve anomaly judgment.
SmartGuard~\cite{zhang2025smartguard} summarizes provenance subgraphs with GNNs and fine-tunes LLMs to better interpret these summaries for detection.
Zuo \emph{et al.}~\cite{zuo2503knowledge} generate natural-language explanations for system events to enrich behavioral signals.

\noindent \textbf{Attack Investigation.}
Beyond initial threat detection, practical security operations demand comprehensive attack reconstruction and forensic analysis capabilities. WATSON~\cite{zeng2021watson} and Sentient~\cite{yan2025sentient} correlate low-level alerts into higher-level scenarios via graph traversal and clustering. UIScope~\cite{yang2020uiscope} augments provenance with GUI events to recover user-interface-driven attacks, while OmegaLog~\cite{hassan2020omegalog} and ALchemist~\cite{yu2021alchemist} fuse system provenance with application logs for richer semantics. ProTracker~\cite{liu2018towards}, NoDoze~\cite{hassan2019nodoze}, Swift~\cite{hassan2020we}, DEPIMPACT~\cite{fang2022back}, and ORTHRUS~\cite{jiang2025orthrus} score events using causal context and aggregate scores over neighboring edges to reconstruct attack graphs. OCR-APT~\cite{aly2025ocr} is an early attempt to use LLMs for investigation, but limits the model to a non-authoritative advisory role in a rigid workflow, constraining autonomous inquiry and collaborative reasoning.

\noindent \textbf{LLMs in Cybersecurity.}
Large language models (LLMs) have demonstrated significant potential in cybersecurity applications~\cite{zhang2025llms,liupropertygpt,tang2025polar,zhang2025attackg+,wen2025vul,gao2025malguard}. They have been widely adopted across downstream tasks such as threat intelligence analysis, vulnerability management, malware detection, and automated code repair. For example, PropertyGPT~\cite{liupropertygpt} generates property specifications for smart-contract formal verification. Polar~\cite{tang2025polar} uses LLMs to prioritize threats by converting unstructured intelligence into structured severity assessments and actionable defenses. MalGuard~\cite{gao2025malguard} semantically refines graph-extracted sensitive APIs to enable interpretable detection of malicious PyPI packages.

\section{Conclusion}

Advanced Persistent Threats (APTs) continue to pose severe challenges to modern cybersecurity, calling for systems that balance efficiency, accuracy, and investigative depth. This paper presents ProvAgent, a threat investigation framework that moves beyond traditional human–model collaboration by integrating detection models with a multi-agent investigation pipeline. ProvAgent enforces fine-grained identity–behavior consistency via graph contrastive learning to perform cost-effective anomaly screening and generate high-fidelity alerts with few false positives. Building on these alerts, the MAI module conducts autonomous, hypothesis–verification–driven investigation to validate and expand initial leads while reconstructing near-complete attack narratives.
Comprehensive evaluations on DARPA E3, E5, and OpTC show that ProvAgent outperforms six SOTA baselines in anomaly detection and remains robust against mimicry attacks. It supports automated, low-cost investigation with a minimum daily cost of \$0.06 and, on DARPA E3, expands the detected IOC set by 160.7\% over initial alerts while producing concise, analyst-oriented reports covering key APT stages and IOCs.

\cleardoublepage
\appendix
\section*{Ethical Considerations}
To the best of our knowledge this work does not raise any ethical issues. All experiments have been performed on publicly available datasets that have been acquired in an ethical manners and not contain any sensitive information.

\section*{Open Science}
\noindent \textbf{Datasets Availability.} The DARPA datasets are publicly available~\cite{darpa_e3,darpa_e5,darpa_optc} and come with textual description of the attacks.

\noindent \textbf{Code Availability.} The code of ProvAgent is publicly available at \url{https://github.com/Win7ery/ProvAgent}.
\bibliographystyle{unsrt}
\bibliography{usenix, \jobname}  

\begin{thebibliography}{10}

\bibitem{alshamrani2019aptsurvey}
Adel Alshamrani, Sowmya Myneni, Ankur Chowdhary, and Dijiang Huang.
\newblock A survey on advanced persistent threats: Techniques, solutions, challenges, and research opportunities.
\newblock {\em IEEE Communications Surveys \& Tutorials}, 21(2):1851--1877, 2019.

\bibitem{li2021threat}
Zhenyuan Li, Qi~Alfred Chen, Runqing Yang, Yan Chen, and Wei Ruan.
\newblock Threat detection and investigation with system-level provenance graphs: A survey.
\newblock {\em Computers \& Security}, 106:102282, 2021.

\bibitem{zipperle2022provenance}
Michael Zipperle, Florian Gottwalt, Elizabeth Chang, and Tharam Dillon.
\newblock Provenance-based intrusion detection systems: A survey.
\newblock {\em ACM Computing Surveys}, 55(7):1--36, 2022.

\bibitem{dong2023we}
Feng Dong, Shaofei Li, Peng Jiang, Ding Li, Haoyu Wang, Liangyi Huang, Xusheng Xiao, Jiedong Chen, Xiapu Luo, Yao Guo, et~al.
\newblock Are we there yet? an industrial viewpoint on provenance-based endpoint detection and response tools.
\newblock In {\em Proceedings of the 2023 ACM SIGSAC Conference on Computer and Communications Security}, pages 2396--2410, 2023.

\bibitem{inam2023sok}
Muhammad~Adil Inam, Yinfang Chen, Akul Goyal, Jason Liu, Jaron Mink, Noor Michael, Sneha Gaur, Adam Bates, and Wajih~Ul Hassan.
\newblock Sok: History is a vast early warning system: Auditing the provenance of system intrusions.
\newblock In {\em 2023 IEEE Symposium on Security and Privacy (SP)}, pages 2620--2638. IEEE, 2023.

\bibitem{xu2025deep}
Zhiwei Xu, Yujuan Wu, Shiheng Wang, Jiabao Gao, Tian Qiu, Ziqi Wang, Hai Wan, and Xibin Zhao.
\newblock Deep learning-based intrusion detection systems: A survey.
\newblock {\em arXiv preprint arXiv:2504.07839}, 2025.

\bibitem{hossain2017sleuth}
Md~Nahid Hossain, Sadegh~M Milajerdi, Junao Wang, Birhanu Eshete, Rigel Gjomemo, R~Sekar, Scott Stoller, and VN~Venkatakrishnan.
\newblock $\{$SLEUTH$\}$: Real-time attack scenario reconstruction from $\{$COTS$\}$ audit data.
\newblock In {\em 26th USENIX Security Symposium (USENIX Security 17)}, pages 487--504, 2017.

\bibitem{milajerdi2019holmes}
Sadegh~M Milajerdi, Rigel Gjomemo, Birhanu Eshete, Ramachandran Sekar, and VN~Venkatakrishnan.
\newblock Holmes: real-time apt detection through correlation of suspicious information flows.
\newblock In {\em 2019 IEEE Symposium on Security and Privacy (SP)}, pages 1137--1152. IEEE, 2019.

\bibitem{hassan2020tactical}
Wajih~Ul Hassan, Adam Bates, and Daniel Marino.
\newblock Tactical provenance analysis for endpoint detection and response systems.
\newblock In {\em 2020 IEEE Symposium on Security and Privacy (SP)}, pages 1172--1189. IEEE, 2020.

\bibitem{xiong2020conan}
Chunlin Xiong, Tiantian Zhu, Weihao Dong, Linqi Ruan, Runqing Yang, Yueqiang Cheng, Yan Chen, Shuai Cheng, and Xutong Chen.
\newblock Conan: A practical real-time apt detection system with high accuracy and efficiency.
\newblock {\em IEEE Transactions on Dependable and Secure Computing}, 19(1):551--565, 2020.

\bibitem{hossain2020combating}
Md~Nahid Hossain, Sanaz Sheikhi, and R~Sekar.
\newblock Combating dependence explosion in forensic analysis using alternative tag propagation semantics.
\newblock In {\em 2020 IEEE Symposium on Security and Privacy (SP)}, pages 1139--1155. IEEE, 2020.

\bibitem{wang2025incorporating}
Lingzhi Wang, Xiangmin Shen, Weijian Li, Zhenyuan Li, R~Sekar, Han Liu, and Yan Chen.
\newblock Incorporating gradients to rules: Towards lightweight, adaptive provenance-based intrusion detection.
\newblock In {\em NDSS}, 2025.

\bibitem{akbar2022advanced}
Khandakar~Ashrafi Akbar, Yigong Wang, Gbadebo Ayoade, Yang Gao, Anoop Singhal, Latifur Khan, Bhavani Thuraisingham, and Kangkook Jee.
\newblock Advanced persistent threat detection using data provenance and metric learning.
\newblock {\em IEEE Transactions on Dependable and Secure Computing}, 20(5):3957--3969, 2022.

\bibitem{kapoor2021prov}
Maya Kapoor, Joshua Melton, Michael Ridenhour, Siddharth Krishnan, and Thomas Moyer.
\newblock Prov-gem: Automated provenance analysis framework using graph embeddings.
\newblock In {\em 2021 20th IEEE International Conference on Machine Learning and Applications (ICMLA)}, pages 1720--1727. IEEE, 2021.

\bibitem{log2vec}
Fucheng Liu, Yu~Wen, Dongxue Zhang, Xihe Jiang, Xinyu Xing, and Dan Meng.
\newblock Log2vec: A heterogeneous graph embedding based approach for detecting cyber threats within enterprise.
\newblock In {\em Proceedings of the 2019 ACM SIGSAC Conference on Computer and Communications Security}, CCS '19, page 1777–1794, New York, NY, USA, 2019. Association for Computing Machinery.

\bibitem{deeplog}
Min Du, Feifei Li, Guineng Zheng, and Vivek Srikumar.
\newblock Deeplog: Anomaly detection and diagnosis from system logs through deep learning.
\newblock In {\em Proceedings of the 2017 ACM SIGSAC Conference on Computer and Communications Security}, CCS '17, page 1285–1298, New York, NY, USA, 2017. Association for Computing Machinery.

\bibitem{ding2023airtag}
Hailun Ding, Juan Zhai, Yuhong Nan, and Shiqing Ma.
\newblock $\{$AIRTAG$\}$: Towards automated attack investigation by unsupervised learning with log texts.
\newblock In {\em 32nd USENIX Security Symposium (USENIX Security 23)}, pages 373--390, 2023.

\bibitem{manzoor2016fast}
Emaad Manzoor, Sadegh~M Milajerdi, and Leman Akoglu.
\newblock Fast memory-efficient anomaly detection in streaming heterogeneous graphs.
\newblock In {\em Proceedings of the 22nd ACM SIGKDD international conference on knowledge discovery and data mining}, pages 1035--1044, 2016.

\bibitem{han2020unicorn}
Xueyuan Han, Thomas Pasquier, Adam Bates, James Mickens, and Margo Seltzer.
\newblock Unicorn: Runtime provenance-based detector for advanced persistent threats.
\newblock In {\em Network and Distributed System Security Symposium}, 2020.

\bibitem{wang2022threatrace}
Su~Wang, Zhiliang Wang, Tao Zhou, Hongbin Sun, Xia Yin, Dongqi Han, Han Zhang, Xingang Shi, and Jiahai Yang.
\newblock Threatrace: Detecting and tracing host-based threats in node level through provenance graph learning.
\newblock {\em IEEE Transactions on Information Forensics and Security}, 17:3972--3987, 2022.

\bibitem{yang2023prographer}
Fan Yang, Jiacen Xu, Chunlin Xiong, Zhou Li, and Kehuan Zhang.
\newblock $\{$PROGRAPHER$\}$: An anomaly detection system based on provenance graph embedding.
\newblock In {\em 32nd USENIX Security Symposium (USENIX Security 23)}, pages 4355--4372, 2023.

\bibitem{rehman2024flash}
Mati~Ur Rehman, Hadi Ahmadi, and Wajih~Ul Hassan.
\newblock Flash: A comprehensive approach to intrusion detection via provenance graph representation learning.
\newblock In {\em 2024 IEEE Symposium on Security and Privacy (SP)}, pages 139--139. IEEE Computer Society, 2024.

\bibitem{cheng2024kairos}
Zijun Cheng, Qiujian Lv, Jinyuan Liang, Yan Wang, Degang Sun, Thomas Pasquier, and Xueyuan Han.
\newblock Kairos: Practical intrusion detection and investigation using whole-system provenance.
\newblock In {\em 2024 IEEE Symposium on Security and Privacy (SP)}, pages 3533--3551. IEEE, 2024.

\bibitem{jia2024magic}
Zian Jia, Yun Xiong, Yuhong Nan, Yao Zhang, Jinjing Zhao, and Mi~Wen.
\newblock $\{$MAGIC$\}$: Detecting advanced persistent threats via masked graph representation learning.
\newblock In {\em 33rd USENIX Security Symposium (USENIX Security 24)}, pages 5197--5214, 2024.

\bibitem{goyal2024r}
Akul Goyal, Gang Wang, and Adam Bates.
\newblock R-caid: Embedding root cause analysis within provenance-based intrusion detection.
\newblock In {\em 2024 IEEE Symposium on Security and Privacy (SP)}, pages 3515--3532. IEEE, 2024.

\bibitem{wu2025brewing}
Weiheng Wu, Wei Qiao, Wenhao Yan, Bo~Jiang, Yuling Liu, Baoxu Liu, Zhigang Lu, and Junrong Liu.
\newblock Brewing vodka: Distilling pure knowledge for lightweight threat detection in audit logs.
\newblock In {\em Proceedings of the ACM on Web Conference 2025}, pages 2172--2182, 2025.

\bibitem{qiao2025slot}
Wei Qiao, Yebo Feng, Teng Li, Zhuo Ma, Yulong Shen, JianFeng Ma, and Yang Liu.
\newblock Slot: Provenance-driven apt detection through graph reinforcement learning.
\newblock In {\em Proceedings of the 2025 on ACM SIGSAC Conference on Computer and Communications Security}, 2025.

\bibitem{bilot2025sometimes}
Tristan Bilot, Baoxiang Jiang, Zefeng Li, Nour El~Madhoun, Khaldoun Al~Agha, Anis Zouaoui, and Thomas Pasquier.
\newblock Sometimes simpler is better: A comprehensive analysis of $\{$State-of-the-Art$\}$$\{$Provenance-Based$\}$ intrusion detection systems.
\newblock In {\em 34th USENIX Security Symposium (USENIX Security 25)}, pages 7193--7212, 2025.

\bibitem{song2024audit}
Chengyu Song, Linru Ma, Jianming Zheng, Jinzhi Liao, Hongyu Kuang, and Lin Yang.
\newblock Audit-llm: Multi-agent collaboration for log-based insider threat detection.
\newblock {\em arXiv preprint arXiv:2408.08902}, 2024.

\bibitem{mukherjee2025llm}
Kunal Mukherjee and Murat Kantarcioglu.
\newblock Llm-driven provenance forensics for threat investigation and detection.
\newblock {\em arXiv preprint arXiv:2508.21323}, 2025.

\bibitem{cheng2025omnisec}
Wenrui Cheng, Tiantian Zhu, Shunan Jing, Jian-Ping Mei, Mingjun Ma, Jiaobo Jin, and Zhengqiu Weng.
\newblock Omnisec: Llm-driven provenance-based intrusion detection via retrieval-augmented behavior prompting.
\newblock {\em Available at SSRN 5397852}, 2025.

\bibitem{zhang2025smartguard}
Hao Zhang, Shuo Shao, Song Li, Zhenyu Zhong, Yan Liu, Zhan Qin, and Kui Ren.
\newblock Smartguard: Leveraging large language models for network attack detection through audit log analysis and summarization.
\newblock {\em arXiv preprint arXiv:2506.16981}, 2025.

\bibitem{zuo2503knowledge}
F~Zuo, J~Rhee, and YR~Choe.
\newblock Knowledge transfer from llms to provenance analysis: A semantic-augmented method for apt detection. arxiv 2025.
\newblock {\em arXiv preprint arXiv:2503.18316}, 2025.

\bibitem{soc_study}
Bushra~A Alahmadi, Louise Axon, and Ivan Martinovic.
\newblock 99\% false positives: A qualitative study of $\{$SOC$\}$ analysts' perspectives on security alarms.
\newblock In {\em 31st USENIX Security Symposium (USENIX Security 22)}, pages 2783--2800, 2022.

\bibitem{alsaheel2021atlas}
Abdulellah Alsaheel, Yuhong Nan, Shiqing Ma, Le~Yu, Gregory Walkup, Z~Berkay Celik, Xiangyu Zhang, and Dongyan Xu.
\newblock $\{$ATLAS$\}$: A sequence-based learning approach for attack investigation.
\newblock In {\em 30th USENIX security symposium (USENIX security 21)}, pages 3005--3022, 2021.

\bibitem{zeng2021watson}
Jun Zeng, Zheng~Leong Chua, Yinfang Chen, Kaihang Ji, Zhenkai Liang, and Jian Mao.
\newblock Watson: Abstracting behaviors from audit logs via aggregation of contextual semantics.
\newblock In {\em NDSS}, 2021.

\bibitem{yan2025sentient}
Wenhao Yan, Ning An, Wei Qiao, Weiheng Wu, Bo~Jiang, Zhigang Lu, and Junrong Liu.
\newblock Sentient: Detecting apts via capturing indirect dependencies and behavioral logic.
\newblock In {\em Proceedings of the AAAI conference on artificial intelligence}, 2026.

\bibitem{aly2025ocr}
Ahmed Aly, Essam Mansour, and Amr Youssef.
\newblock Ocr-apt: Reconstructing apt stories from audit logs using subgraph anomaly detection and llms.
\newblock In {\em Proceedings of the 2025 ACM SIGSAC Conference on Computer and Communications Security}, pages 261--275, 2025.

\bibitem{darpa_e3}
A.~D. Keromytis.
\newblock Transparent computing engagement 3 data release, 2018.
\newblock Accessed 24th December 2025. \url{https://github.com/darpa-i2o/Transparent-Computing/blob/master/README-E3.md}.

\bibitem{darpa_e5}
J.~Torrey.
\newblock Transparent computing engagement 5 data release, 2018.
\newblock Accessed 24th December 2025. \url{https://github.com/darpa-i2o/Transparent-Computing}.

\bibitem{darpa_optc}
Mike van Opstal and William Arbaugh.
\newblock Operationally transparent cyber (optc) data release, 2019.
\newblock Accessed 24th December 2025. \url{https://github.com/FiveDirections/OpTC-data}.

\bibitem{zengy2022shadewatcher}
Jun Zengy, Xiang Wang, Jiahao Liu, Yinfang Chen, Zhenkai Liang, Tat-Seng Chua, and Zheng~Leong Chua.
\newblock Shadewatcher: Recommendation-guided cyber threat analysis using system audit records.
\newblock In {\em 2022 IEEE Symposium on Security and Privacy (SP)}, pages 489--506. IEEE, 2022.

\bibitem{jiang2025orthrus}
Baoxiang Jiang, Tristan Bilot, Nour El~Madhoun, Khaldoun Al~Agha, Anis Zouaoui, Shahrear Iqbal, Xueyuan Han, and Thomas Pasquier.
\newblock Orthrus: Achieving high quality of attribution in provenance-based intrusion detection systems.
\newblock In {\em Security Symposium (USENIX Sec’25). USENIX}, 2025.

\bibitem{ullah2024llms}
Saad Ullah, Mingji Han, Saurabh Pujar, Hammond Pearce, Ayse Coskun, and Gianluca Stringhini.
\newblock Llms cannot reliably identify and reason about security vulnerabilities (yet?): A comprehensive evaluation, framework, and benchmarks.
\newblock In {\em 2024 IEEE symposium on security and privacy (SP)}, pages 862--880. IEEE, 2024.

\bibitem{zibaeirad2025reasoning}
Arastoo Zibaeirad and Marco Vieira.
\newblock Reasoning with llms for zero-shot vulnerability detection.
\newblock {\em arXiv preprint arXiv:2503.17885}, 2025.

\bibitem{habibzadeh2025large}
Ali Habibzadeh, Farid Feyzi, and Reza~Ebrahimi Atani.
\newblock Large language models for security operations centers: A comprehensive survey.
\newblock {\em arXiv preprint arXiv:2509.10858}, 2025.

\bibitem{linux_audit_daemon}
{Linux Documentation}.
\newblock The linux audit daemon.
\newblock Accessed 24th December 2025. \url{https://man7.org/linux/man-pages/man8/auditd.8.html}.

\bibitem{windows}
Microsoft.
\newblock Event tracing.
\newblock Accessed 24th December 2025. \url{https://learn.microsoft.com/en-us/windows-hardware/drivers/devtest/about-event-tracing-for-drivers}.

\bibitem{camflow}
Thomas Pasquier, Xueyuan Han, Mark Goldstein, Thomas Moyer, David Eyers, Margo Seltzer, and Jean Bacon.
\newblock Practical whole-system provenance capture.
\newblock In {\em Proceedings of the 2017 Symposium on Cloud Computing}, pages 405--418, 2017.

\bibitem{sekar2024eaudit}
R~Sekar, Hanke Kimm, and Rohit Aich.
\newblock eaudit: A fast, scalable and deployable audit data collection system.
\newblock In {\em 2024 IEEE Symposium on Security and Privacy (SP)}, pages 3571--3589. IEEE, 2024.

\bibitem{mikolov2013efficient}
Tomas Mikolov.
\newblock Efficient estimation of word representations in vector space.
\newblock {\em arXiv preprint arXiv:1301.3781}, 2013.

\bibitem{ju2024towards}
Wei Ju, Yifan Wang, Yifang Qin, Zhengyang Mao, Zhiping Xiao, Junyu Luo, Junwei Yang, Yiyang Gu, Dongjie Wang, Qingqing Long, et~al.
\newblock Towards graph contrastive learning: A survey and beyond.
\newblock {\em arXiv preprint arXiv:2405.11868}, 2024.

\bibitem{ocr_apt_code}
Ocr-apt code.
\newblock Accessed: 24th December 2025. \url{https://github.com/CoDS-GCS/OCR-APT}.

\bibitem{darpa2018tc3_groundtruth}
{DARPA}.
\newblock {TC3 Ground Truth Report}, 2018.
\newblock Accessed: 24th December 2025. \url{https://drive.google.com/file/d/1mrs4LWkGk-3zA7t7v8zrhm0yEDHe57QU/view}.

\bibitem{goyal2023sometimes}
Akul Goyal, Xueyuan Han, Gang Wang, and Adam Bates.
\newblock Sometimes, you aren’t what you do: Mimicry attacks against provenance graph host intrusion detection systems.
\newblock In {\em 30th Network and Distributed System Security Symposium}, 2023.

\bibitem{mitre_attack}
{MITRE ATT\&CK}.
\newblock \url{https://attack.mitre.org}, 2019.

\bibitem{hamilton2017inductive}
Will Hamilton, Zhitao Ying, and Jure Leskovec.
\newblock Inductive representation learning on large graphs.
\newblock {\em Advances in neural information processing systems}, 30, 2017.

\bibitem{vaswani2017attention}
A~Vaswani.
\newblock Attention is all you need.
\newblock {\em Advances in Neural Information Processing Systems}, 2017.

\bibitem{rossi2020temporal}
Emanuele Rossi, Ben Chamberlain, Fabrizio Frasca, Davide Eynard, Federico Monti, and Michael Bronstein.
\newblock Temporal graph networks for deep learning on dynamic graphs.
\newblock In {\em International Conference on Machine Learning (ICML’20)}, 2020.

\bibitem{darvariu2024graph}
Victor-Alexandru Darvariu, Stephen Hailes, and Mirco Musolesi.
\newblock Graph reinforcement learning for combinatorial optimization: A survey and unifying perspective.
\newblock {\em arXiv preprint arXiv:2404.06492}, 2024.

\bibitem{yang2020uiscope}
Runqing Yang, Shiqing Ma, Haitao Xu, Xiangyu Zhang, and Yan Chen.
\newblock Uiscope: Accurate, instrumentation-free, and visible attack investigation for gui applications.
\newblock In {\em NDSS}, volume~24, page 141, 2020.

\bibitem{hassan2020omegalog}
Wajih~Ul Hassan, Mohammad~Ali Noureddine, Pubali Datta, and Adam Bates.
\newblock Omegalog: High-fidelity attack investigation via transparent multi-layer log analysis.
\newblock In {\em Network and distributed system security symposium}, 2020.

\bibitem{yu2021alchemist}
Le~Yu, Shiqing Ma, Zhuo Zhang, Guanhong Tao, Xiangyu Zhang, Dongyan Xu, Vincent~E Urias, Han~Wei Lin, Gabriela~F Ciocarlie, Vinod Yegneswaran, et~al.
\newblock Alchemist: Fusing application and audit logs for precise attack provenance without instrumentation.
\newblock In {\em NDSS}, 2021.

\bibitem{liu2018towards}
Yushan Liu, Mu~Zhang, Ding Li, Kangkook Jee, Zhichun Li, Zhenyu Wu, Junghwan Rhee, and Prateek Mittal.
\newblock Towards a timely causality analysis for enterprise security.
\newblock In {\em NDSS}, 2018.

\bibitem{hassan2019nodoze}
Wajih~Ul Hassan, Shengjian Guo, Ding Li, Zhengzhang Chen, Kangkook Jee, Zhichun Li, and Adam Bates.
\newblock Nodoze: Combatting threat alert fatigue with automated provenance triage.
\newblock In {\em network and distributed systems security symposium}, 2019.

\bibitem{hassan2020we}
Wajih~Ul Hassan, Ding Li, Kangkook Jee, Xiao Yu, Kexuan Zou, Dawei Wang, Zhengzhang Chen, Zhichun Li, Junghwan Rhee, Jiaping Gui, et~al.
\newblock This is why we can’t cache nice things: Lightning-fast threat hunting using suspicion-based hierarchical storage.
\newblock In {\em Proceedings of the 36th Annual Computer Security Applications Conference}, pages 165--178, 2020.

\bibitem{fang2022back}
Pengcheng Fang, Peng Gao, Changlin Liu, Erman Ayday, Kangkook Jee, Ting Wang, Yanfang~Fanny Ye, Zhuotao Liu, and Xusheng Xiao.
\newblock $\{$Back-Propagating$\}$ system dependency impact for attack investigation.
\newblock In {\em 31st USENIX security symposium (USENIX Security 22)}, pages 2461--2478, 2022.

\bibitem{zhang2025llms}
Jie Zhang, Haoyu Bu, Hui Wen, Yongji Liu, Haiqiang Fei, Rongrong Xi, Lun Li, Yun Yang, Hongsong Zhu, and Dan Meng.
\newblock When llms meet cybersecurity: A systematic literature review.
\newblock {\em Cybersecurity}, 8(1):55, 2025.

\bibitem{liupropertygpt}
Ye~Liu, Yue Xue, Daoyuan Wu, Yuqiang Sun, Yi~Li, Miaolei Shi, and Yang Liu.
\newblock Propertygpt: Llm-driven formal verification of smart contracts through retrieval-augmented property generation.
\newblock In {\em NDSS}, 2025.

\bibitem{tang2025polar}
Luoxi Tang, Yuqiao Meng, Ankita Patra, Weicheng Ma, Muchao Ye, and Zhaohan Xi.
\newblock Polar: Automating cyber threat prioritization through llm-powered assessment.
\newblock {\em arXiv preprint arXiv:2510.01552}, 2025.

\bibitem{zhang2025attackg+}
Yongheng Zhang, Tingwen Du, Yunshan Ma, Xiang Wang, Yi~Xie, Guozheng Yang, Yuliang Lu, and Ee-Chien Chang.
\newblock Attackg+: Boosting attack graph construction with large language models.
\newblock {\em Computers \& Security}, 150:104220, 2025.

\bibitem{wen2025vul}
Xin-Cheng Wen, Zirui Lin, Yijun Yang, Cuiyun Gao, and Deheng Ye.
\newblock Vul-r2: A reasoning llm for automated vulnerability repair.
\newblock In {\em Proceedings of the 40th IEEE/ACM International Conference on Automated Software Engineering (ASE)}, 2025.

\bibitem{gao2025malguard}
Xingan Gao, Xiaobing Sun, Sicong Cao, Kaifeng Huang, Di~Wu, Xiaolei Liu, Xingwei Lin, and Yang Xiang.
\newblock Malguard: Towards real-time, accurate, and actionable detection of malicious packages in pypi ecosystem.
\newblock In {\em Proceedings of the 34th USENIX Security Symposium (USENIX Security 25)}, 2025.

\end{thebibliography}

\section{Identity–Behavior Binding Performance}

\begin{figure}
    \centering
    \includegraphics[width=\linewidth]{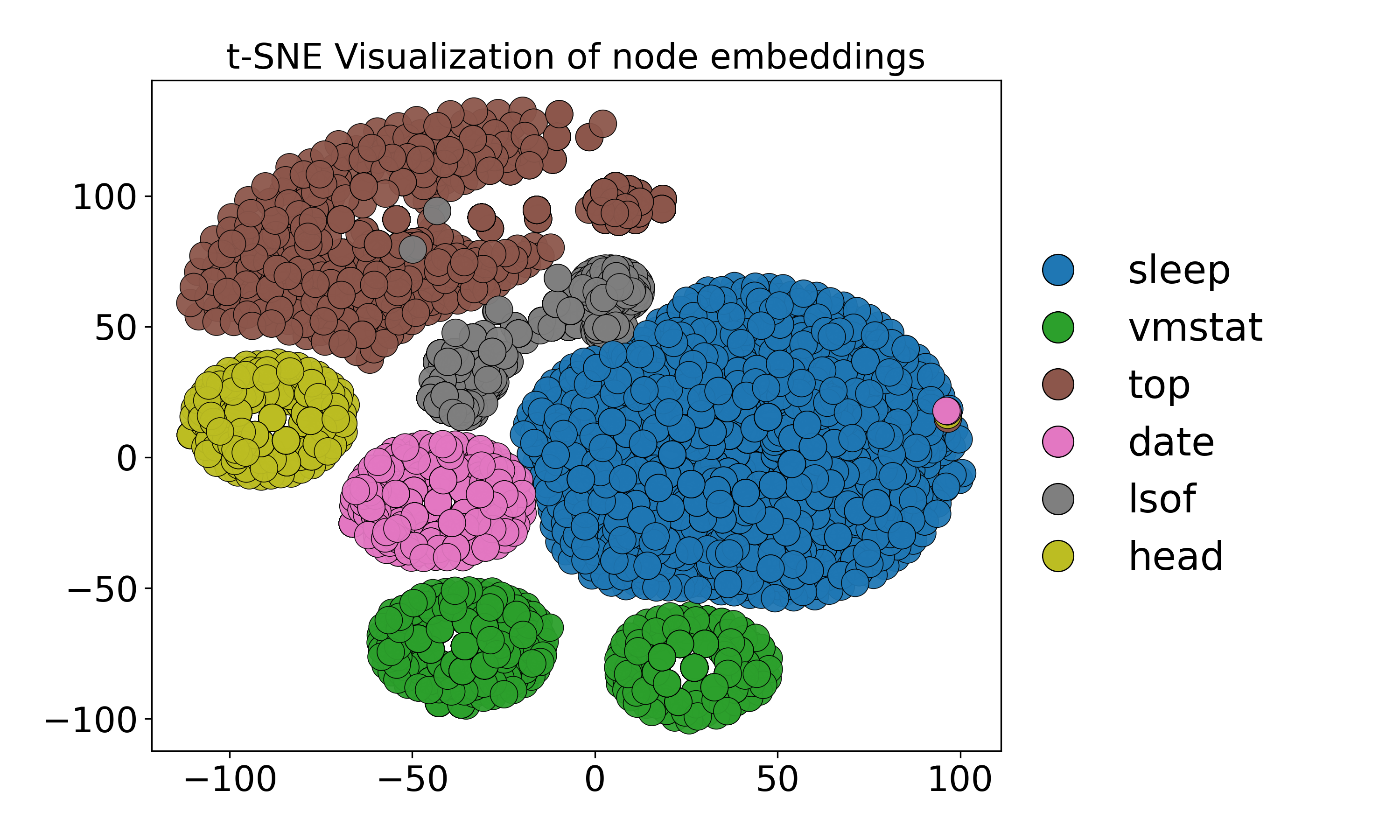}
    \caption{T-SNE visualization of entity embeddings on DARPA-E3. Each node is a system entity and is colored by identity. The legend shows the six most frequent process identities, including \texttt{sleep}, \texttt{vmstat}, \texttt{top}, \texttt{date}, \texttt{lsof}, and \texttt{head}.}
    \label{fig:tsne}
\end{figure}

To validate that ProvAgent's EPD module successfully binds entity identity with behavioral patterns in the learned embedding space, we visualize the latent representations using t-SNE on the DARPA-E3 dataset. Figure~\ref{fig:tsne} shows the two-dimensional projection of entity embeddings, where each point represents a system entity and is colored according to its fine-grained identity (as defined in Table~\ref{tab:identity}).

The visualization demonstrates clear clustering by identity, with entities sharing the same identity forming tight, well-separated clusters in the embedding space. For instance, processes with identities \texttt{sleep}, \texttt{vmstat}, \texttt{top}, \texttt{date}, \texttt{lsof}, and \texttt{head} each occupy distinct regions, indicating that the contrastive learning framework effectively enforces identity-specific behavioral consistency. The clear boundaries between different identity clusters validate our hypothesis that entities with identical identities exhibit similar behavioral patterns, and that these patterns can be reliably captured through graph neural network encoding combined with metric learning.

This clustering property enables ProvAgent's identity-consistency violation detection mechanism: anomalous entities whose behaviors deviate from their claimed identity will be projected into regions inconsistent with their declared identity cluster, thereby facilitating interpretable anomaly detection.

\section{LLM Prompts}

This section presents the system prompts designed for the four specialized agents in the MAI module to define their roles and operational guidelines. Due to space constraints, we provide only the core system instructions herein while the complete prompt templates are available in our online repository~\footnote{\url{https://github.com/Win7ery/ProvAgent}}.

\subsection{Leader Agent Prompt}

The Leader agent performs strategic-level coordination and global coherence validation. It synthesizes fragmented behaviors into coherent attack narratives by mapping them onto structured kill chain frameworks, generates strategic hypotheses when gaps or inconsistencies emerge, and determines investigation termination criteria based on narrative coherence and hypothesis exhaustion.

\noindent\fbox{\parbox{0.96\linewidth}{
\small
\textbf{Leader Agent System Prompt:} You are the Leader of the Security Operations Center. You read the IOC table for a single graph, align each IOC with the cyber kill chain, and direct further investigation when gaps exist.

Responsibilities:

1. Connect the dots across IOCs to form a cohesive attack story.

2. Use each IOC's \texttt{event} description plus \texttt{related\_ioc\_*} 

3. fields to explain relationships.

4. Mark which kill-chain phases are satisfied (with evidence) and which remain missing.

5. Request additional investigations by referencing concrete \texttt{node\_uuid} targets.

6. Flag false positives so they can be removed from the IOC store.

7. Always output structured JSON exactly as requested.
}}

\subsection{Analyst Agent Prompt}

The Analyst agent validates whether alerts or suspicious leads represent genuine anomalies by comparing target entities against benign behavioral baselines retrieved from the knowledge repository. It conducts fine-grained behavioral comparison using both attribute-based string matching and embedding-based vector similarity to discriminate genuine malicious deviations from benign operational heterogeneity.

\noindent\fbox{\parbox{0.96\linewidth}{
\small
\textbf{Analyst Agent System Prompt:} You are the Analyst. Every input (alert or clue) is a single node that might be malicious. Compare it against trustworthy context to decide whether it is ANOMALY or BENIGN.

Guidance:

1. Run two comparisons: FAISS most-similar node and label-matched benign peers. Inconsistency between the target node's behavior and its similar node does NOT automatically indicate anomaly. Only flag as anomalous when the inconsistent behavior itself exhibits anomalous characteristics.

2. Treat \texttt{investigator\_reason} as a hint only; stay objective.

3. Cite concrete neighbor edges, operations, or temporal behaviors.

4. Output \texttt{confidence} between 0 and 1 even when BENIGN.
}}

\subsection{Investigator Agent Prompt}

The Investigator agent conducts event-level forensic analysis to reconstruct atomic attack operations from confirmed IOCs. It traverses temporal event sequences and provenance neighborhoods, identifies related neighboring entities as new investigative leads, and annotates each IOC with kill chain stage mappings and detailed event descriptions.

\noindent\fbox{\parbox{0.96\linewidth}{
\small
\textbf{Investigator Agent System Prompt:} You are the Investigator. Each task focuses on a confirmed IOC. Validate the node's malicious behavior, assign the correct kill-chain phase (reason), narrate the attack, and propose new clues for the Analyst. Produce an \texttt{event} summary that captures the concrete malicious action. Every clue you output must include \texttt{graph\_name}, \texttt{node\_uuid}, and \texttt{reason} so the Analyst queries the correct graph. Just describe why they look risky.
}}

\subsection{Reporter Agent Prompt}

The Reporter agent transforms the investigation repository into analyst-ready reports. It organizes validated IOCs into coherent attack narratives, constructs provenance graphs grounded in event evidence, and summarizes actionable remediation guidance for security analysts.

\noindent\fbox{\parbox{0.96\linewidth}{
\small
\textbf{Reporter Agent System Prompt:} You are the Reporter. Use structured attack data to produce a high-quality Markdown report for human security analysts. Highlight the campaign flow, major IOCs, and rationale for each IOC's malicious verdict.
}}

\section{Attack Reconstruction}\label{appendix:attack_reconstruction}

This section presents the attack provenance graphs automatically reconstructed by ProvAgent for five APT campaigns in the DARPA E3 dataset. These visualizations demonstrate ProvAgent's capability to produce compact, analyst-oriented attack narratives without human intervention. Each graph is generated by the Reporter agent based on the validated IOCs and enriched metadata accumulated during the multi-agent investigation process.

\begin{figure}
    \centering
    \includegraphics[width=\linewidth]{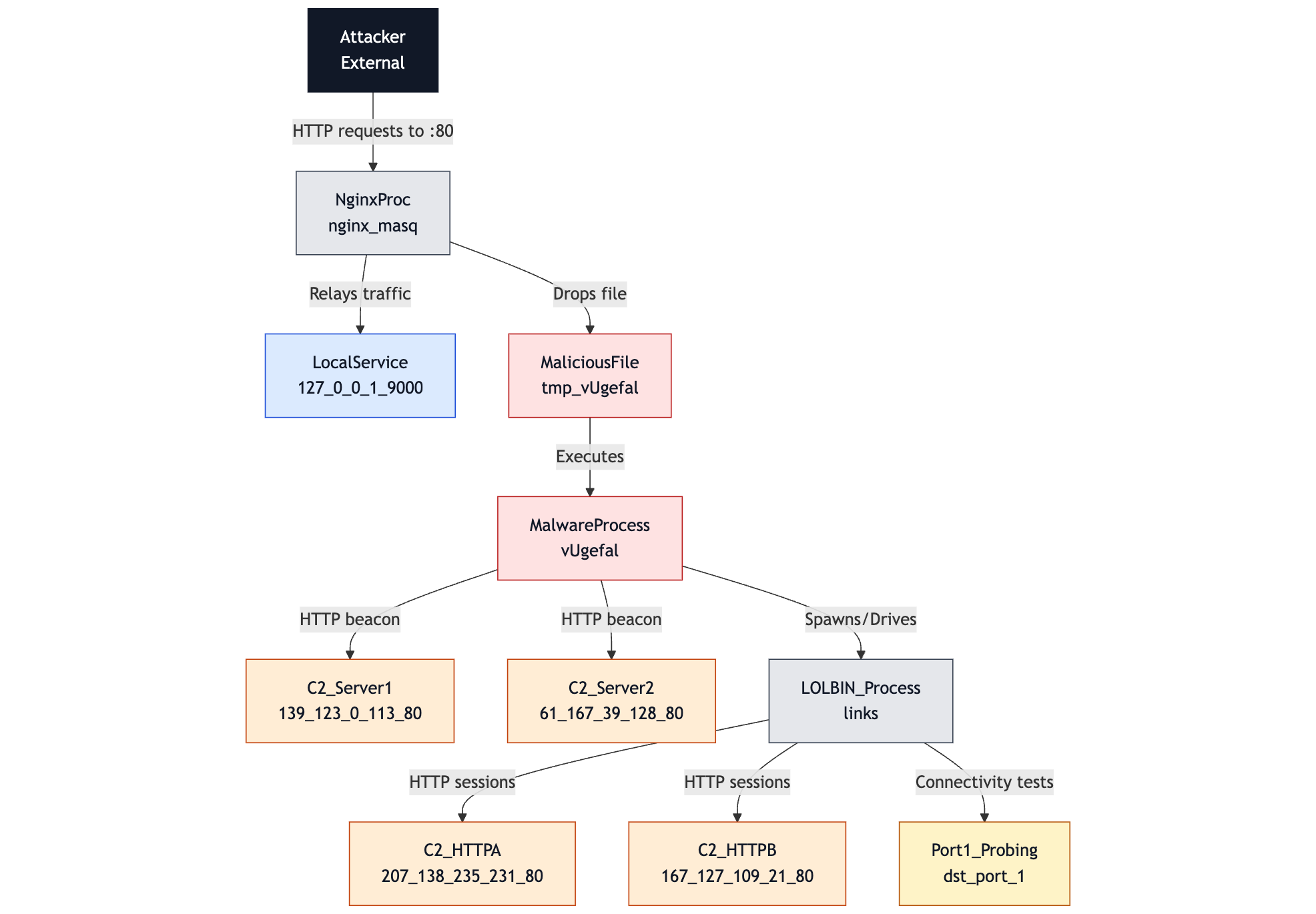}
    \caption{Attack graph automatically generated by ProvAgent for the April 6 attack on CADETS-E3.}
    \label{fig:cadets6}
\end{figure}

\begin{figure}
    \centering
    \includegraphics[width=\linewidth]{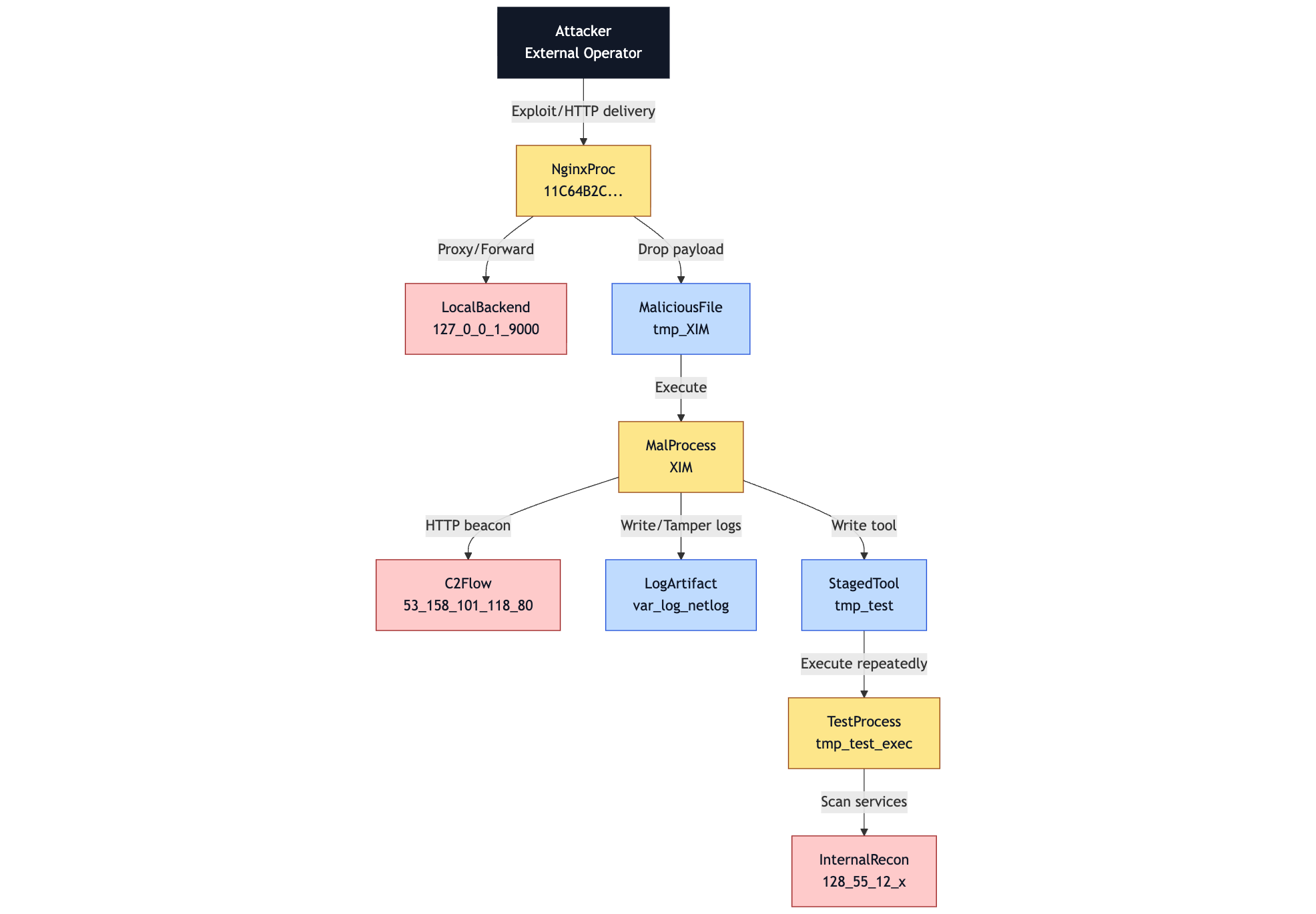}
    \caption{Attack graph automatically generated by ProvAgent for the April 12 attack on CADETS-E3.}
    \label{fig:cadets12}
\end{figure}

\begin{figure}
    \centering
    \includegraphics[width=\linewidth]{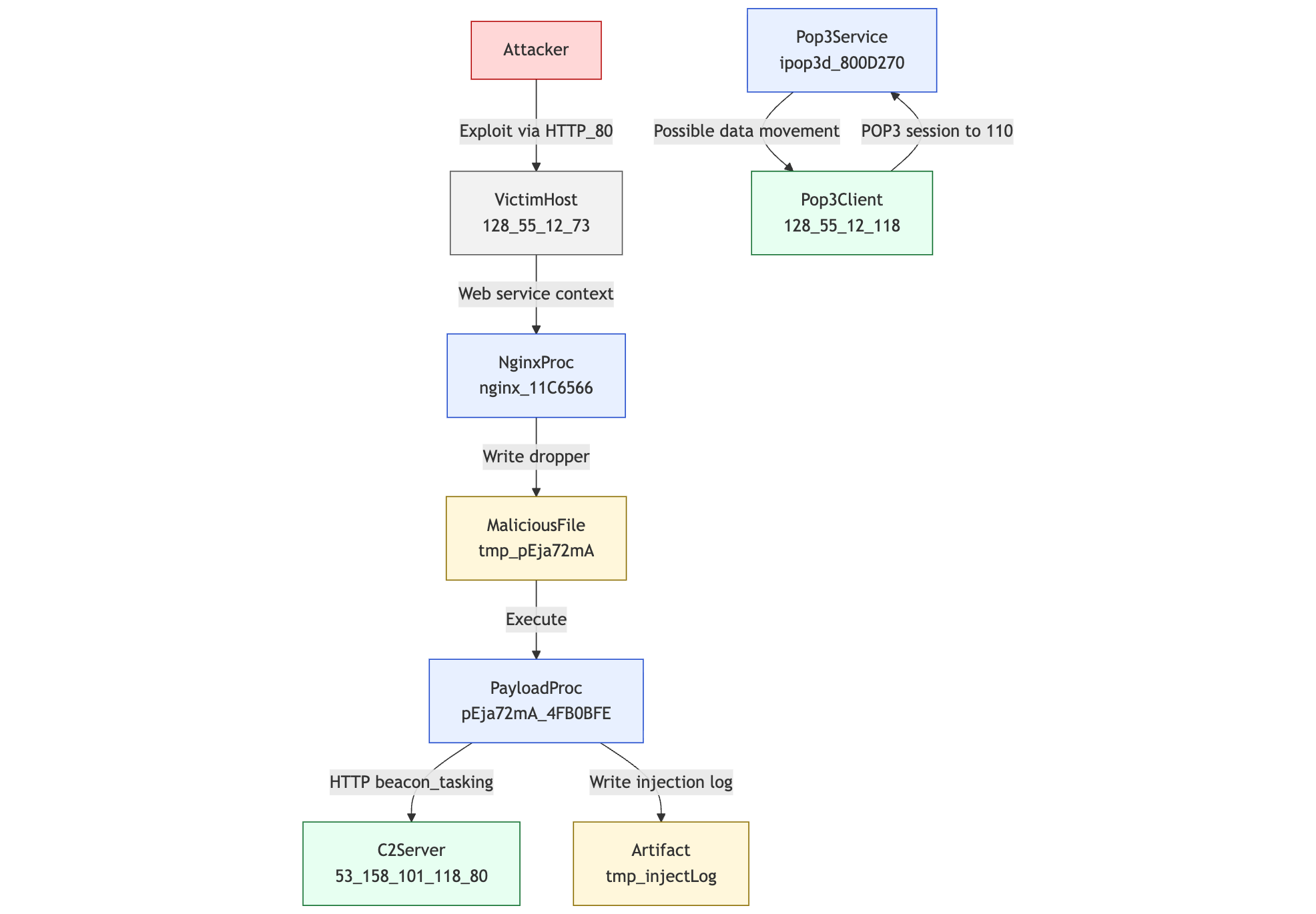}
    \caption{Attack graph automatically generated by ProvAgent for the April 13 attack on CADETS-E3.}
    \label{fig:cadets13}
\end{figure}

\begin{figure}
    \centering
    \includegraphics[width=\linewidth]{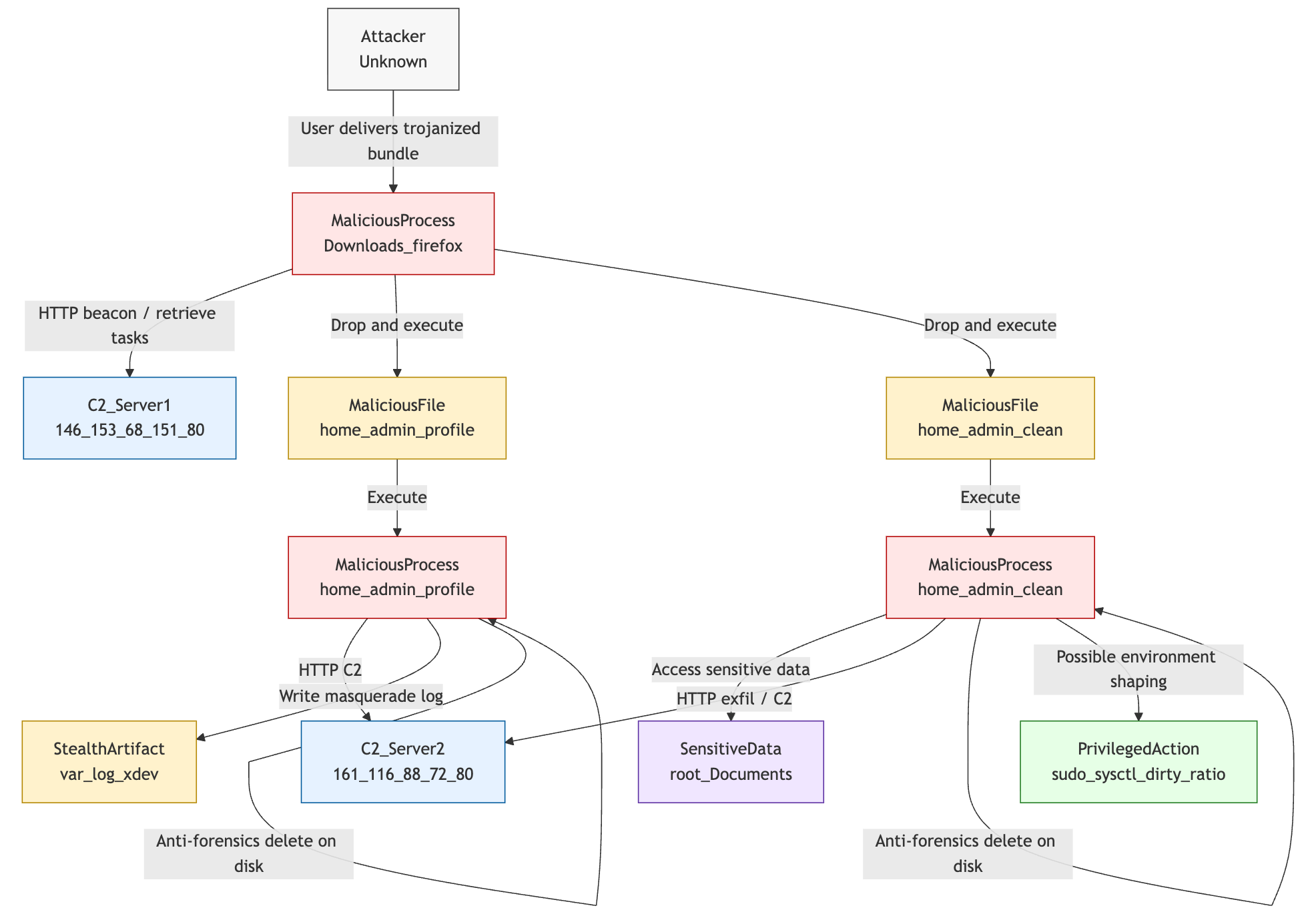}
    \caption{Attack graph automatically generated by ProvAgent for the April 10 attack on THEIA-E3.}
    \label{fig:theia10}
\end{figure}

\begin{figure}
    \centering
    \includegraphics[width=\linewidth]{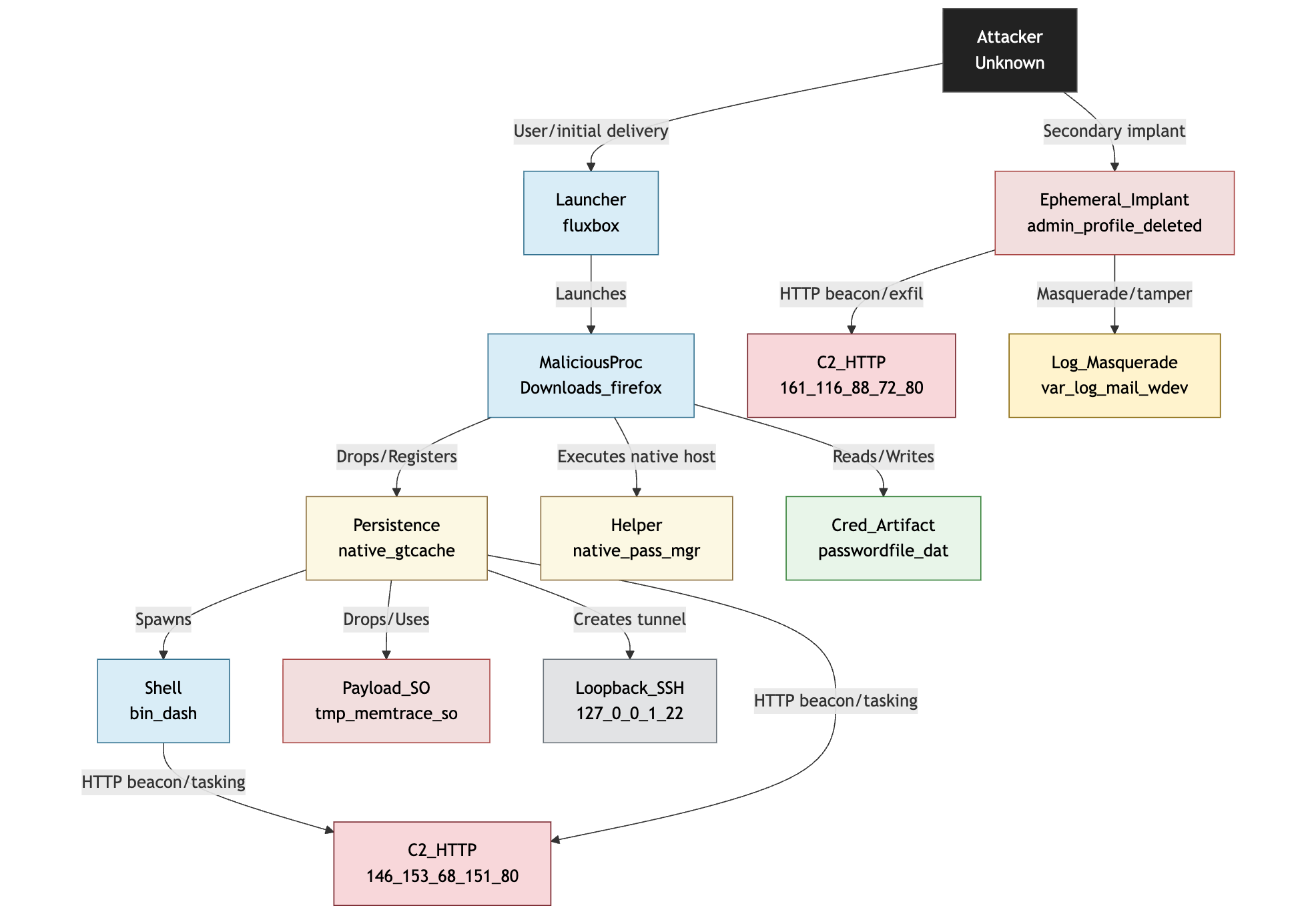}
    \caption{Attack graph automatically generated by ProvAgent for the April 12 attack on THEIA-E3.}
    \label{fig:theia12}
\end{figure}

Figures~\ref{fig:cadets6}, \ref{fig:cadets12}, and \ref{fig:cadets13} show the reconstructed attack graphs for the nginx backdoor campaigns on April 6, April 12, and April 13 in the CADETS-E3 dataset. These attacks involve web server compromise, command-and-control communication, privilege escalation, and data manipulation. Figures~\ref{fig:theia10} and \ref{fig:theia12} present the reconstructed graphs for the Firefox backdoor and browser extension attacks on April 10 and April 12 in the THEIA-E3 dataset, which involve initial compromise through browser exploitation, internal reconnaissance, persistence mechanisms, and credential theft.

\section{Analysis of Similar Nodes}\label{appendix:similar}

Although ProvAgent substantially reduces the false-positive rate compared to prior baselines, our analysis shows that most remaining false positives stem from behavioral similarity between processes with different identities. While the EPD module enforces identity–behavior consistency via contrastive learning, some system entities serve distinct functional roles yet exhibit highly similar syscall-level patterns.

\begin{figure}[htbp]
    \centering
    \begin{subfigure}[b]{0.45\columnwidth}
        \centering
        \includegraphics[width=\textwidth]{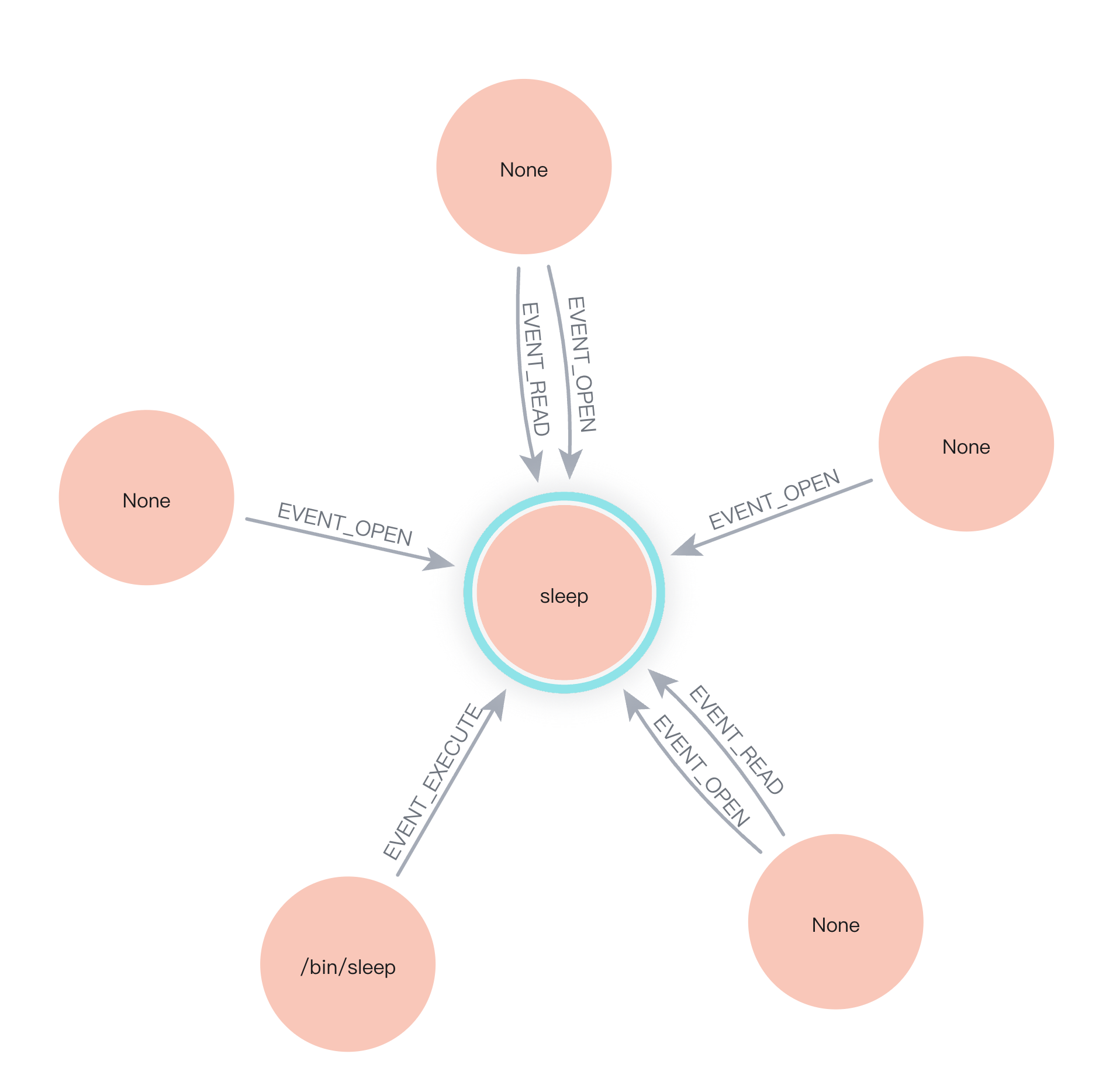}
        \caption{The behavior graph of the sleep process.}
        \label{fig:sub1}
    \end{subfigure}
    \hfill
    \begin{subfigure}[b]{0.45\columnwidth}
        \centering
        \includegraphics[width=\textwidth]{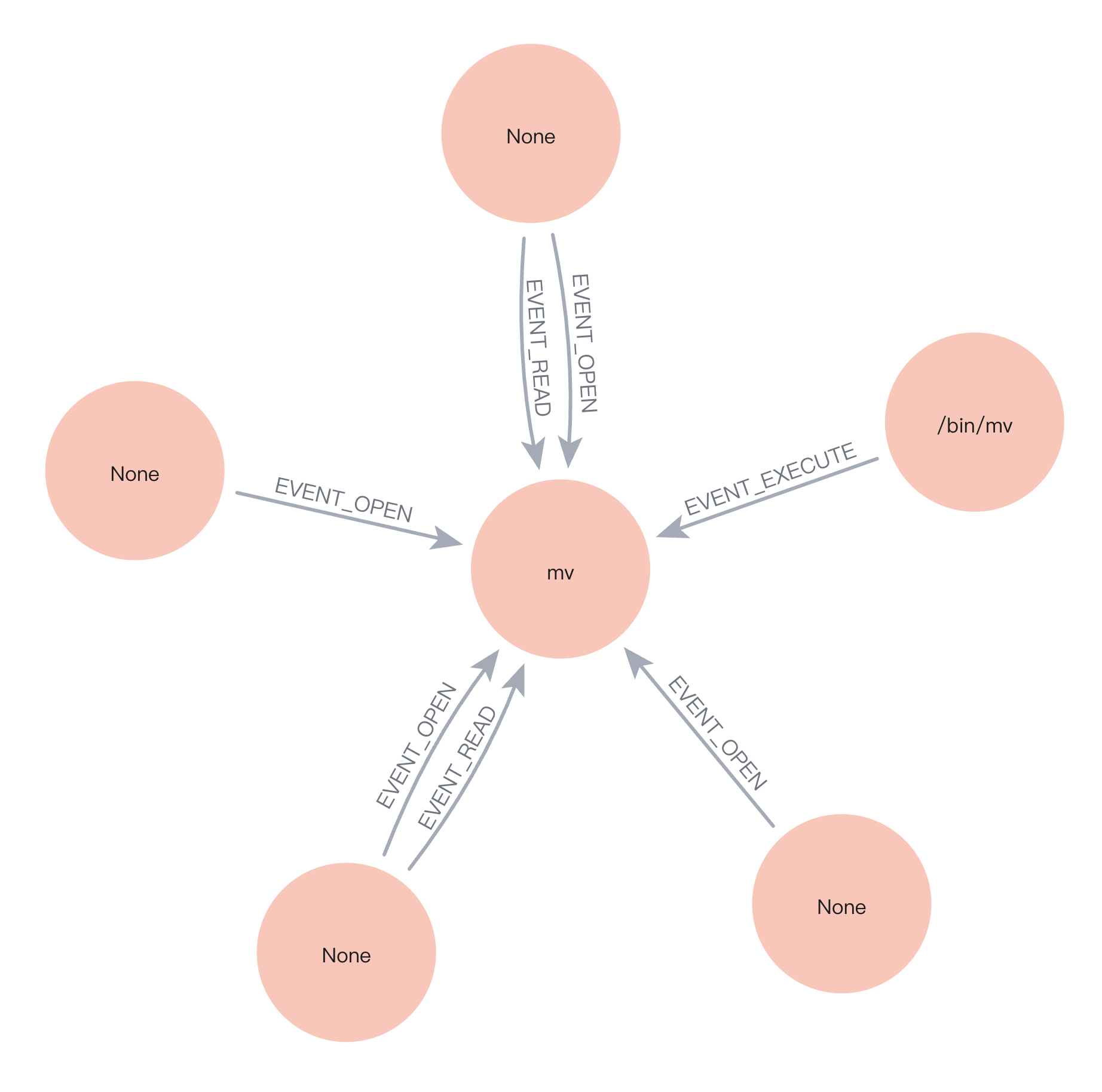}
        \caption{The behavior graph of the mv process.}
        \label{fig:sub2}
    \end{subfigure}
    
    \begin{subfigure}[b]{0.45\columnwidth}
        \centering
        \includegraphics[width=\textwidth]{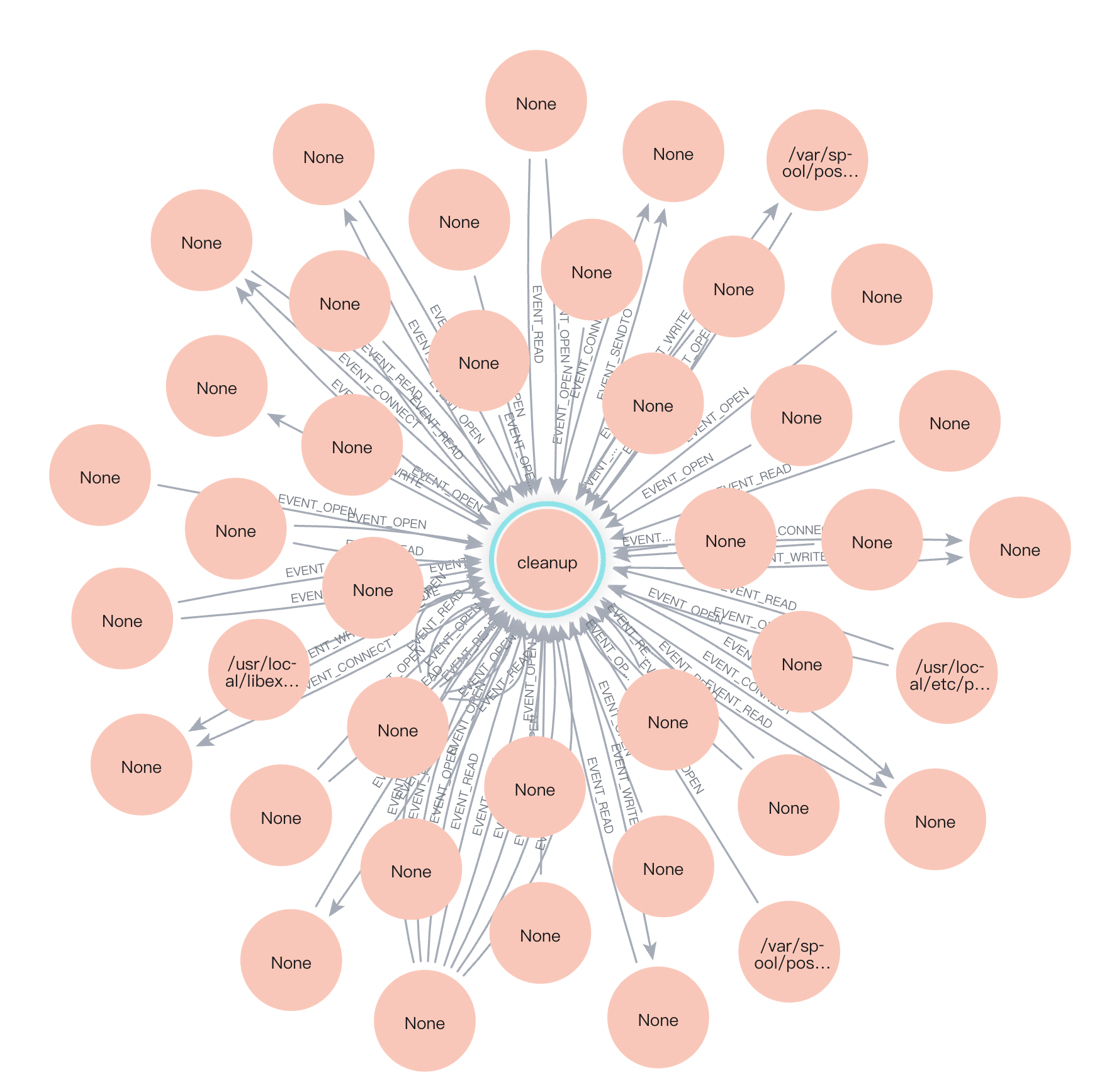}
        \caption{The behavior graph of the cleanup process.}
        \label{fig:sub3}
    \end{subfigure}
    \hfill
    \begin{subfigure}[b]{0.45\columnwidth}
        \centering
        \includegraphics[width=\textwidth]{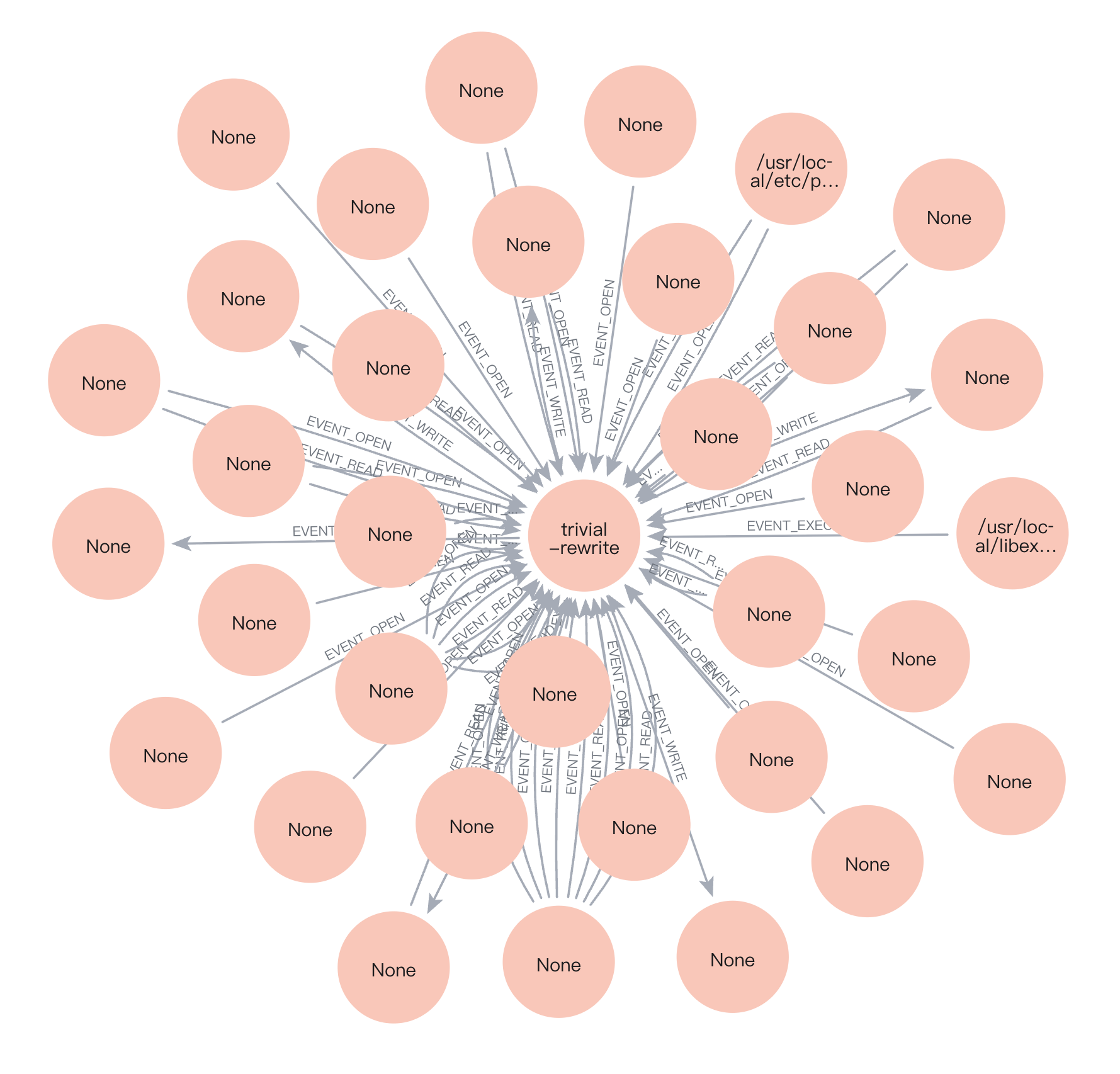}
        \caption{The behavior graph of the trivial-rewrite process.}
        \label{fig:sub4}
    \end{subfigure}
    
    \caption{Examples of similar node pairs causing false positives in ProvAgent. The first pair includes (a) \texttt{sleep} and (b) \texttt{mv} process, while the second pair consists of (c) \texttt{cleanup} and (d) \texttt{trivial-rewrite} process. Each subfigure shows the behavioral sequences of the process.}
    \label{fig:overall}
\end{figure}

Figure~\ref{fig:overall} illustrates two representative process pairs that frequently cause identity confusion in ProvAgent. The first pair, \texttt{sleep} and \texttt{mv}, consists of short-lived utility processes that perform only a few file-descriptor operations and share nearly identical syscall sequences. The second pair, \texttt{cleanup} and \texttt{trivial-rewrite}, both interact heavily with \texttt{none} files and access \texttt{/usr/local/libexec/postfix/*}. In both cases, the behavioral patterns captured by the GNN are not sufficiently discriminative, causing EPD to align entities with the wrong identity prototypes in the embedding space.

This limitation is mainly due to the coarse granularity of syscall-level observations. Although ProvAgent’s Word2Vec encoding captures semantic cues from process names and file paths, frequency and temporal signals are largely dominated by low-level syscall patterns and may fail to reflect higher-level functional semantics. As a result, processes with similar syscall sequences, regardless of their functional intent, tend to cluster closely in the learned embedding space, leading to potential identity mismatches during anomaly detection.

\end{document}